\documentclass[amssymb,amsmath,aps,secnumarabic,floatfix,twocolumn,pre,showpacs,showkeys,superscriptaddress,balancelastpage]{revtex4}

\usepackage{graphicx}
\usepackage{xspace}
\usepackage{subfigure}
\usepackage{longtable}
\usepackage{subfigure}

\newcommand{\latin}[1]{{\it #1}}
\newcommand{\ie}{\latin{i.e.}\@\xspace}
\newcommand{\eg}{\latin{e.g.}\@\xspace}

\makeatletter
\newcommand{\supmarker}[1]{{\@ifempty{#1}{}{\text{(#1)}}}}
\makeatother
\newcommand{\pdf}[2]{\mathcal{P}^{(#1)}\left(#2\right)}

\newcommand{\slabel}[1]{\label{sec:#1}}

\newcommand{\Sref}[1]{Sec.~\ref{sec:#1}}

\newcommand{\elabel}[1]{\label{eq:#1}}
\newcommand{\eref}[1]{Eq.~(\ref{eq:#1})}

\newcommand{\Eref}[1]{Eq.~(\ref{eq:#1})}

\newcommand{\flabel}[1]{\label{fig:#1}}

\newcommand{\Fref}[1]{Fig.~\ref{fig:#1}}
\newcommand{\Freft}[2]{Figs.~\ref{fig:#1} and \ref{fig:#2}}

\newcommand{\tlabel}[1]{\label{table:#1}}

\newcommand{\Tref}[1]{Table~\ref{table:#1}}
\newcommand{\Trefs}[2]{Tables~\ref{table:#1} and \ref{table:#2}}

\newcommand{\GC}{\mathcal{G}}

\newcommand{\plaind}{\mathrm{d}}
\newcommand{\dint}[1]{\mathchoice{\!\plaind#1\,}{\!\plaind#1\,}{\!\plaind#1\,}{\!\plaind#1\,}}

\newcommand{\ave}[1]{\left\langle #1 \right\rangle}
\newcommand{\cov}[1]{\text{cov}(#1)}
\newcommand{\covx}[2]{\cov{x^{#1},x^{#2}}}

\newcommand{\nFUTA}{\nFUTAshort{} lattice\xspace}
\newcommand{\nFUTAshort}{Futatsubishi\xspace}
\newcommand{\FUTA}{\nFUTA{} (\Fref{futatsubishi})\xspace}
\newcommand{\nLADD}{\nLADDshort}
\newcommand{\nLADDshort}{rope ladder\xspace}

\newcommand{\nLINE}{\nLINEshort}
\newcommand{\nLINEshort}{simple chain\xspace}
\newcommand{\LINE}{\nLINE{} (\Fref{line})\xspace}
\newcommand{\nNNN}{\nNNNshort}
\newcommand{\nNNNshort}{nnn chain\xspace} 
\newcommand{\NNN}{\nNNN{} (\Fref{nnn})\xspace}
\newcommand{\nARCH}{\nARCHshort{} lattice\xspace} 
\newcommand{\nARCHshort}{Archimedes\xspace} 
\newcommand{\ARCH}{\nARCH{} (\Fref{archimedes})\xspace}
\newcommand{\nHONE}{\nHONEshort{} lattice\xspace} 
\newcommand{\nHONEshort}{honeycomb\xspace} 
\newcommand{\HONE}{\nHONE{} (\Fref{honeycomb})\xspace}
\newcommand{\nJAGG}{\nJAGGshort{} lattice\xspace} 
\newcommand{\nJAGGshort}{jagged\xspace} 
\newcommand{\JAGG}{\nJAGG{} (\Fref{jaggy})\xspace}
\newcommand{\nKAGO}{\nKAGOshort{} lattice\xspace} 
\newcommand{\nKAGOshort}{Kagom{\'e}\xspace} 
\newcommand{\KAGO}{\nKAGO{} (\Fref{kagome})\xspace}
\newcommand{\nMITS}{\nMITSshort{} lattice\xspace} 
\newcommand{\nMITSshort}{Mitsubishi\xspace} 
\newcommand{\MITS}{\nMITS{} (\Fref{mitsubishi})\xspace}
\newcommand{\nNOCR}{\nNOCRshort{} lattice\xspace} 
\newcommand{\nNOCRshort}{nc diagonal square\xspace} 
\newcommand{\NOCR}{\nNOCR{} (\Fref{nocrossing})\xspace}
\newcommand{\nSQUA}{\nSQUAshort{} lattice\xspace} 
\newcommand{\nSQUAshort}{square\xspace} 
\newcommand{\SQUA}{\nSQUA{} (\Fref{square})\xspace}
\newcommand{\nTRIA}{\nTRIAshort{} lattice\xspace} 
\newcommand{\nTRIAshort}{triangular\xspace} 
\newcommand{\TRIA}{\nTRIA{} (\Fref{triangular})\xspace}

\newcommand{\numact}{N_a}
\newcommand{\fssExpo}[2]{\mu^{(#1)}_{#2}}
\newcommand{\momrat}[2]{g^{(#1)}_{#2}}
\newcommand{\Btilde}{\tilde{B}}
\newcommand{\qoflow}[1]{[#1]}

\newcommand{\strong}[1]{}

\expandafter\ifx\csname package@font\endcsname\relax\else
  \expandafter\expandafter
  \expandafter\usepackage
  \expandafter\expandafter
  \expandafter{\csname package@font\endcsname}%
\fi

\begin{document}
\title{Abelian Manna model on various lattices in one and two dimensions}
\date{\today}

\author{Hoai Nguyen Huynh}
\email{hu0004en@e.ntu.edu.sg}
\homepage{http://www3.ntu.edu.sg/home2008/hu0004en/}
\affiliation{Division of Physics and Applied Physics,
School of Physical and Mathematical Sciences,
Nanyang Technological University, Singapore,
21 Nanyang Link, Singapore 637371}

\author{Gunnar Pruessner}
\email{g.pruessner@imperial.ac.uk}
\homepage{http://www.ma.ic.ac.uk/~pruess}
\affiliation{Department of Mathematics,
Imperial College London,
180 Queen's Gate,
London SW7 2BZ, United Kingdom}

\author{Lock Yue Chew}
\email{lockyue@ntu.edu.sg}
\homepage{http://www.ntu.edu.sg/home/lockyue/}
\affiliation{Division of Physics and Applied Physics,
School of Physical and Mathematical Sciences,
Nanyang Technological University, Singapore,
21 Nanyang Link, Singapore 637371}

\begin{abstract}
We perform a high-accuracy moment analysis of the avalanche size, duration and
area distribution of the Abelian Manna model on eight two-dimensional and four
one-dimensional lattices.
The results provide strong support to establish universality of exponents and
moment ratios across different lattices and a good survey for the strength of
corrections to scaling which are notorious in the Manna universality class. The
results are compared against previous work done on Manna model, Oslo model and
directed percolation. We also confirm hypothesis of various scaling relations.
\end{abstract}

\pacs{
05.65.+b, 
05.70.Jk 
}

\keywords{Self-organized criticality,
Lattices,
Universality,
Finite-size scaling,
Scaling relations,
Moments}

\maketitle

\section{Introduction}
\slabel{introduction}
Universality is the key feature of critical systems which justifies the analysis of
(over-) simplified numerical models of otherwise much more complex natural
systems. When the concept of self-organised criticality (SOC) was  introduced
\cite{BakTangWiesenfeld:1987}, it gained immediate popularity on the one hand because it
attempted to explain the prevalence of scaling and fractality in nature and on the
other hand suggested that many features of very different phenomena would be
common to all of them, by the power of universality. At times disputed
\cite{ChristensenOlami:1992a}, it is now widely accepted that the predictive
power of SOC lies with its universality; the elusive qualities of a critical
point, normally confined to a very narrow range of a control parameter, are the
norm in SOC models and shared between them.

While universality has been at the centre of the debate about SOC
\cite[\eg][]{NakanishiSneppen:1997,Ben-HurBiham:1996}, it comes as a great
surprise that it has focused mainly on its occurrence across different models
\cite[\eg][]{Dhar:2006}, whereas very little work has been done on establishing it
within the same models but across different lattices
\cite[\eg][]{Duarte:1990,Manna:1991b,HuLin:2003,Azimi-TafreshiETAL:2010}.
Recently, it has been shown
\cite{HuynhChewPruessner:2010} that on fractal lattices of same dimension but different
structure, the Manna Model \cite{Manna:1991a} behaves very differently. The observation that critical
phenomena and scaling displays universal features on different lattices has
traditionally been one of the most important insights, enabling, in particular, exact
results \cite{Syozi:1951,ItzyksonDrouffe:1997}.

During the last decade or so, it has become increasingly clear that SOC
is far more elusive than originally envisaged. It is therefore all the
more important to establish which features known from ordinary critical
phenomena carry over to SOC. Independence of certain observables from the
underlying lattice is one such aspect to substantiate.  If universality
was not to be found for the same model on different lattices, then the
notion of universality in SOC as a whole had to be revised.

In the following, the universality hypothesis in SOC is put to test by
simulating the Abelian Manna model (AMM) \cite{Manna:1991a,Dhar:1999a} on
different one- and two-dimensional lattices.  Owing to their robust
scaling behaviour, in recent years, attention has focused on 
the AMM and the Oslo Model \cite{ChristensenETAL:1996}, which seem to be
in the same universality class \cite{NakanishiSneppen:1997}. This is contrasted by the poor
scaling behaviour of many other established SOC models, such as the Forest Fire
Model \cite{JensenPruessner:2002b,Grassberger:2002a}, the
Bak-Tang-Wiesenfeld sandpile model \cite{BakTangWiesenfeld:1987,DornHughesChristensen:2001} or the
Olami-Feder-Christensen Model
\cite{OlamiFederChristensen:1992,ChristensenOlami:1992a}.

Apart from probing universality, corrections to scaling can also
be tested, with the ultimate aim of finding a lattice that is particularly suited for
simulating the AMM. In addition, rarely studied moment ratios are also
reported which provide further support for the universality of the AMM.
This is of particular interest in 
one-dimension, where (logarithmic)
corrections, are known to be remarkably
strong \cite{DickmanCampelo:2003}.

The following is divided into four parts: In the next section, the model is defined
and a few implementation details are discussed, including the method employed to
determine the statistical error. All numerical results are collected in \Sref{results}
and discussed in \Sref{discussion}.

\section{Model definition and methods}
\slabel{model}
\subsection{The Abelian Manna model on an arbitrary lattice}
The Abelian Manna model \cite{Manna:1991a,Dhar:1999a,Dhar:1999c} 
is defined on a lattice $\mathcal{L}$ with $N$ sites. Each site $i$ on
$\mathcal{L}$ has $q_i$ neighbours, and is
assigned to it a non-negative integer variable $z_i$ which can be thought of as the
number of particles at that site or the height of a stack of particles
there. The threshold height at all sites is $1$
above which a site is said to be active or unstable, otherwise it is stable.
The system evolves as follows. \emph{Driving:}
When the system is in a
stable or ``quiescent'' configuration, \ie $z_i\le1$ for all sites, the system is charged by picking a
site $i$ at random and incrementing $z_i$ by 1. \emph{Relaxation:} 
Every unstable site $i$ relaxes by transferring two particles to its
neighbours, possibly rendering the receiving site unstable.
The recipient of each of the two particles is chosen
randomly and independently.
Unstable sites are updated in a random sequential order.
By virtue of its Abelian nature the order of relaxations is irrelevant
for (the statistics of) the final state of the lattice and the
statistics of the avalanche size (see below). 

Each relaxation of a site constitutes a
toppling, which in the bulk is conservative, \ie the total $\sum_i z_i$ remains
unchanged by bulk topplings. A (small) number of sites are considered dissipative
boundary sites, which have ``virtual'' neighbours, that are included in
the total count of neighbours $q_i$ introduced above 
(but not in the number of sites $N$).
Virtual neighbours (or sites) never topple themselves, \ie particles are lost from the system at receipt.
Such sites therefore provide a dissipation mechanism, and are introduced here solely as a bookkeeping device.
 As a general
principle, the number of virtual neighbours of sites in the lattices discussed
below are always chosen so that the resulting $q_i$ at a boundary site
matches that of a corresponding site in the bulk. In the lattices below,
$q_i$ might take on at most two different values, yet bulk sites and
boundary sites do not differ in that respect. 

\begin{table*}[t]
\caption{Parameters for the one-dimensional lattices. Only the seven
sizes eventually used are listed. The (nominal) linear size $L$, corresponding
to the examples given in \Fref{1d_lattices}, is shown in brackets.
\tlabel{system_sizes_1D}
}
\begin{tabular}{ll|lllllll|l}
\hline\hline
lattice & $d$  &
 & & & $N$ & & & & CPU\\
&&&&&&&&&time (h)\\
\hline
Simple chain\xspace	& 1  & $1024 (1024)\ $  & $2048 (2048)\ $  & $4096 (4096)\ $  & $8192 (8192)\ $  & $16384 (16384)\ $  & $32768 (32768)\ $  & $65536 (65536)\ $&	51350 \\
Rope ladder\xspace	& 1  & $2048 (1024)\ $  & $4096 (2048)\ $  & $8192 (4096)\ $  & $16384 (8192)\ $  & $32768 (16384)\ $  & $65536 (32768)\ $  & $131072 (65536)\ $&	54299 \\
\nNNNshort	& 1  & $2048 (1024)\ $  & $4096 (2048)\ $  & $8192 (4096)\ $  & $16384 (8192)\ $  & $32768 (16384)\ $  & $65536 (32768)\ $  & $131072 (65536)\ $&	7943 \\
\nFUTAshort	& 1  & $3073 (1024)\ $  & $6145 (2048)\ $  & $12289 (4096)\ $  & $24577 (8192)\ $  & $49153 (16384)\ $  & $98305 (32768)\ $  & $196609 (65536)\ $&	54821 \\

\hline\hline
\end{tabular}
\end{table*}

An avalanche is the totality of all topplings until the system is
quiescent again.
The size
$s$ of the avalanche is measured as the number topplings performed between driving
and quiescence, 
so that $s=0$ if no avalanche occurs after the driving.
The area $a$ of an avalanche is the
number of distinct sites
which received a particle during the avalanche. This includes 
the site charged by the driving, so that $a\ge1$. The
definition of time used to determine the duration $t$ of an avalanche is based on
the idea that each
active site 
undergoes a Poissonian decay, \ie all active sites topple with the same
rate. 
This rate is chosen to be unity, so that on average time $1/\numact$ goes by until a
site topples if $\numact$ sites are active. This is the amount by which time advances
each time a site topples. Each time a site is picked at random,
exactly two of its particles are redistributed.
This procedure is, of course, not a faithful representation of a true Poisson
process, which has random waiting times, but it is an increasingly
good approximation in the limit of large activity $\numact$ \cite{Liggett:2005}.
Thanks to its Abelian nature \cite{Dhar:1999a}, the model's evolution
from quiescent state to quiescent state does not require a specific
dynamics on the microscopic time scale, so the definition of time in the present case
is to the same degree arbitrary as the dynamics in, say, model A
\cite{HohenbergHalperin:1977}.

The Abelian symmetry does
not exist in the original model \cite{Manna:1991a}, but simplifies its
implementation greatly.
Originally, \emph{all} particles were re-distributed
at toppling which were subject to parallel updates,
so that all sites that were active at time $t$ had
toppled before sites activated by a toppling during that time step. In
that version of the model, it
takes one time unit to update $\numact$ simultaneously active sites. In the Abelian version, this
holds only for constant $\numact$ (number of active sites remaining
unchanged). Yet, the update rate for both versions is $1/\numact$ when averaged
over suitable intervals (one time unit in the former, one update in the
later), although neither is Poissonian.
Despite these differences in the definition, Dickman and
Campelo \cite{DickmanCampelo:2003} have shown that 
implementing parallel or sequential updating has no noticeable
impact on the statistics. They also found that exponents derived from 
the Abelian variant of the Manna Model coincides with those 
published for the original version.

\subsection{The lattices}
In the following, we describe four one-dimensional (\Fref{1d_lattices})
and eight two-dimensional lattices (\Freft{2d_lattices_square}{2d_lattices_hex}). In
the figures, links to adjacent sites are indicated
by solid lines, those to virtual neighbours by dashed ones. As virtual
neighbours are merely an accounting device, the positioning of the dashed
lines in the figures is arbitrarily chosen to match that of bulk links.

\begin{table*}[t]
\caption{Parameters for the two-dimensional lattices. Only the seven
sizes eventually used are listed. The number of sites $N$ equals the product
$L_xL_y$.
\tlabel{system_sizes_2D}
}
\begin{tabular}{ll|llll|lllllll|l}
\hline\hline
lattice & $d$ & $c_x$ & $m_x$ & $c_y$ & $m_y$ &
 & & & $L_x\times L_y$ & & & & CPU\\
&&&&&&&&&&&&&time (h)\\
\hline
Square\xspace	& 2 & 	0 & 1 & 0 & 1       & $256 \!\!\times\!\! 256\ $  & $362 \!\!\times\!\! 362\ $  & $512 \!\!\times\!\! 512\ $  & $724 \!\!\times\!\! 724\ $  & $1024 \!\!\times\!\! 1024\ $  & $1448 \!\!\times\!\! 1448\ $  & $2048 \!\!\times\!\! 2048\ $&	9530 \\
Jagged\xspace	& 2 & 	1 & 2 & 0 & 1       & $361 \!\!\times\!\! 182\ $  & $511 \!\!\times\!\! 257\ $  & $723 \!\!\times\!\! 363\ $  & $1023 \!\!\times\!\! 513\ $  & $1447 \!\!\times\!\! 725\ $  & $2047 \!\!\times\!\! 1025\ $  & $2895 \!\!\times\!\! 1449\ $&	9408 \\
\nARCHshort	& 2 & 	0 & 4 & 0 & 2       & $360 \!\!\times\!\! 182\ $  & $512 \!\!\times\!\! 256\ $  & $724 \!\!\times\!\! 362\ $  & $1024 \!\!\times\!\! 512\ $  & $1448 \!\!\times\!\! 724\ $  & $2048 \!\!\times\!\! 1024\ $  & $2896 \!\!\times\!\! 1448\ $&	11106 \\
\nNOCRshort	& 2 & 	1 & 2 & 1 & 2       & $255 \!\!\times\!\! 257\ $  & $361 \!\!\times\!\! 363\ $  & $511 \!\!\times\!\! 513\ $  & $723 \!\!\times\!\! 725\ $  & $1023 \!\!\times\!\! 1025\ $  & $1447 \!\!\times\!\! 1449\ $  & $2047 \!\!\times\!\! 2049\ $&	9802 \\
Triangular\xspace	& 2 & 	0 & 1 & 0 & 1       & $239 \!\!\times\!\! 274\ $  & $337 \!\!\times\!\! 389\ $  & $476 \!\!\times\!\! 551\ $  & $673 \!\!\times\!\! 779\ $  & $953 \!\!\times\!\! 1100\ $  & $1347 \!\!\times\!\! 1557\ $  & $1906 \!\!\times\!\! 2201\ $&	6305 \\
\nKAGOshort	& 2 & 	1 & 3 & 0 & 1       & $412 \!\!\times\!\! 159\ $  & $583 \!\!\times\!\! 225\ $  & $826 \!\!\times\!\! 317\ $  & $1168 \!\!\times\!\! 449\ $  & $1651 \!\!\times\!\! 635\ $  & $2335 \!\!\times\!\! 898\ $  & $3301 \!\!\times\!\! 1271\ $&	10881 \\
Honeycomb\xspace	& 2 & 	1 & 2 & 0 & 2       & $337 \!\!\times\!\! 194\ $  & $475 \!\!\times\!\! 276\ $  & $675 \!\!\times\!\! 388\ $  & $953 \!\!\times\!\! 550\ $  & $1347 \!\!\times\!\! 778\ $  & $1903 \!\!\times\!\! 1102\ $  & $2695 \!\!\times\!\! 1556\ $&	7842 \\
\nMITSshort	& 2 & 	2 & 3 & 1 & 2       & $239 \!\!\times\!\! 275\ $  & $338 \!\!\times\!\! 387\ $  & $476 \!\!\times\!\! 551\ $  & $674 \!\!\times\!\! 777\ $  & $953 \!\!\times\!\! 1101\ $  & $1349 \!\!\times\!\! 1555\ $  & $1904 \!\!\times\!\! 2203\ $&	7573 \\

\hline\hline
\end{tabular}
\end{table*}

The figures
show in particular the boundaries, whose structure we maintained during
finite size scaling, \ie when we increased the lattice size, we made
sure that the edges and corners of the larger lattice matched that
of the smaller one. Initially we dismissed such issues as small
corrections, but given the high accuracy with which we determine
moments, changes in the boundaries became clearly visible, in particular
in the \MITS. In hindsight this is hardly surprising given the great
importance of transport of particles from the conservative bulk to the
dissipative boundaries \cite{PaczuskiBassler:2000}.

For some simple lattices, the number of sites $N$ is exactly the power
$d$ (dimension) of their linear (Euclidean) extension $L$.  In more complicated
cases, this may hold only approximately. In
order to maintain a reasonably high symmetry of the finite lattices,
which have to be thought of as being cut out of an (infinite) tiling of
the plane, we had to
make compromises. 
In one dimension ($d=1$, see \Tref{system_sizes_1D}), $N$ is a multiple of 
$L$ for the first three lattices discussed. The forth, the \FUTA, follows $N=3L+1$
for all sizes considered. 
In two dimensions, $d=2$, two different lengths $L_x$ and
$L_y$ are introduced for the two different dimensions, defined in a
way as natural as possible.
The cutting of the finite two-dimensional lattices and the 
resulting shape of the boundaries
was guided by the following principles, which clash and therefore hold
only approximately: the number of sites in the lattice $N=L_xL_y$ had to be as
close as possible to the power of $2$, the number of sites in a simple square
lattice.
The ratio $L_x/L_y$ was to be held constant with increasing $N$. The (Euclidean)
aspect ratio should be unity or very close, if that clashes with the shape of the
boundaries. For each lattice $m_{x,y}$ and $c_{x,y}$ were set such that
$L_{x,y} \equiv c_{x,y} \pmod{m_{x,y}}$, to maintain the shape of the boundaries
across different sizes.
The parameters and resulting lattice sizes are listed in \Tref{system_sizes_2D}.
The number $L_y$ can normally be thought as a number of ``layers'' and $L_x$ as
the number of sites in a layer. It is clear that such a definition, necessary because
of the mismatch of the symmetry of lattice to the square symmetry of the cut-out,
makes the linear sizes $L_x$ and $L_y$ rather poor fitting parameters (see below).

The size of each lattice is indicated exemplarily in the caption of its
figure. The sizes of all lattices used in this article are 
listed in \Trefs{system_sizes_1D}{system_sizes_2D}.
In the following we describe the lattices in detail.

\subsubsection{One-dimensional lattices}
Simple chain (\Fref{line}):
This lattice is the usual one-dimensional chain, where each site
connects to two nearest neighbours. The leftmost and the rightmost sites
have one virtual neighbour each.

Rope ladder (\Fref{ladder}):
The shape of this lattice is that of a rope ladder. It is a
simple
extension of the \LINE, which eventually leads to the \SQUA. 
Each site has three nearest
neighbours, one on the left, one on the right and one below or above it. 
Four boundary sites have one virtual neighbour each.

Next nearest neighbour (nnn) chain (\Fref{nnn}):
Despite its triangular pattern, this is the \LINE extended by
allowing for next nearest neighbour interactions. This lattice was
motivated by the observation \cite{HughesPaczuski:2002} that some models
require such extensions in one dimensions to prevent degeneracy.
Alternatively, it can be seen as the first step towards a two-dimensional \TRIA.
Each site has four neighbours, with boundary sites having either one or
two virtual neighbours.

Futatsubishi lattice (\Fref{futatsubishi}):
In appreciation of the Kagom{\'e} lattice discussed below,
``Futatsubishi'' is a Japanese name, which translates to
``two-diamond'', reflecting the shape of the lattice. The Futatsubishi lattice is fully contained in the \MITS
discussed below (three of them meet at every vertex), or as a suitable slice of the \SQUA or the \JAGG.
On the present lattice, each site has either
two or four neighbours in the bulk and two virtual neighbours at the
boundary.

\begin{figure}
\subfigure[The \nLINE. $L=10, N=10$.\flabel{line}]{\includegraphics[scale=0.45]{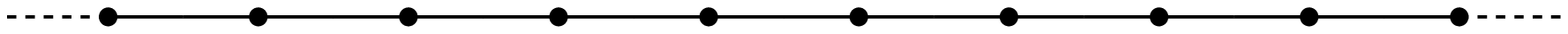}}
\subfigure[The \nLADD. $L=10, N=20$.\flabel{ladder}]{\includegraphics[scale=0.45]{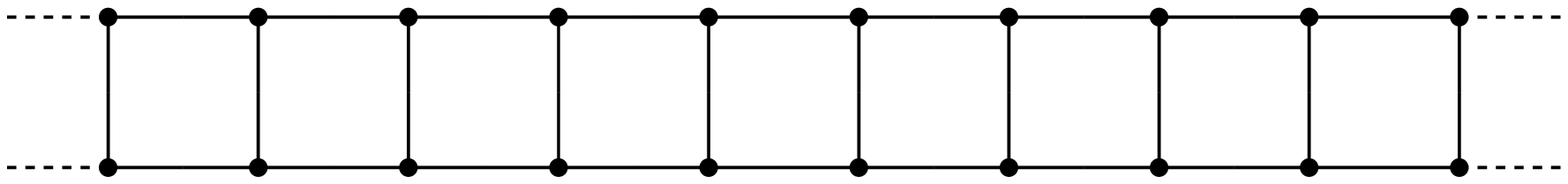}}
\subfigure[The next nearest neighbour (nnn) chain. $L=10, N=20$.\flabel{nnn}]{\includegraphics[scale=0.45]{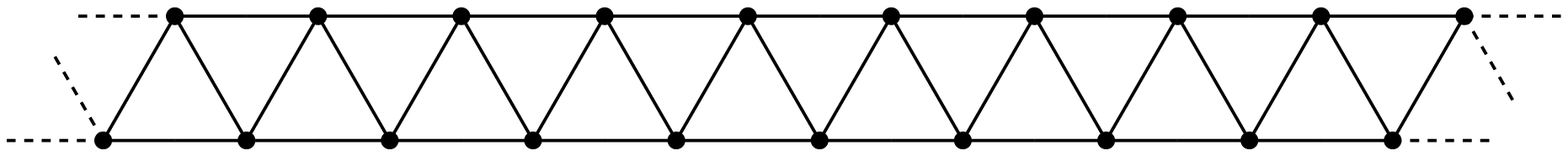}}
\subfigure[The \nFUTA. $L=7, N=22$.\flabel{futatsubishi}]{\includegraphics[scale=0.45]{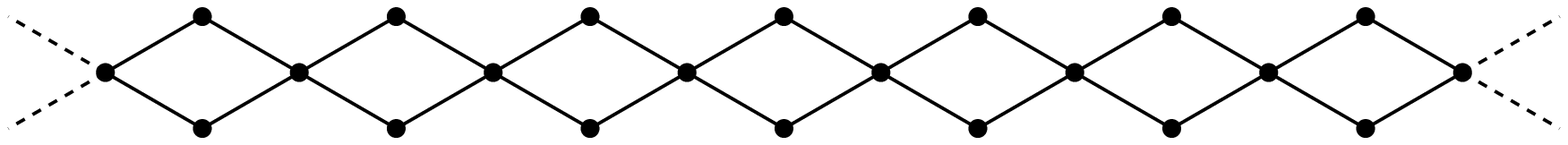}}
\caption{
\flabel{1d_lattices}
The four one-dimensional lattices considered in this article.
Sites are shown as filled circled, adjacency is indicated by solid
lines. Dashed lines indicate links to virtual neighbours. }
\end{figure}

\subsubsection{Two-dimensional lattices}
Square lattice (\Fref{square}): This is the standard square lattice, the most
commonly used lattice
the AMM has been studied on. Each site has four nearest neighbours.
As indicated by the dashed lines in \Fref{square}, edge sites have one
virtual neighbour and corner sites have two, leading to a constant
$q_i=4$ for all sites $i$.

Jagged lattice (\Fref{jaggy}): This lattice is a square lattice rotated by $45$
degrees and
fitted into a square shape which makes the boundary look jagged. The
only difference to the square lattice above is therefore the boundary,
where sites have either two (edge) or three (corner) virtual
neighbours, producing, again, a constant
$q_i=4$ for all sites. There are various ways of cutting the lattice 
out of the bulk --- we decided to maintain the left-right mirror 
symmetry, which results in three different boundaries. In hindsight,
a slightly different choice would have resulted in a lattice of higher
symmetry and an aspect ratio closer to unity. The perfect match of the universal features of the AMM on this lattice
with those on all others is testament of the strong universal behaviour of
the model.

Archimedes lattice (\Fref{archimedes}): The name of this lattice is normally
complemented by $(4,8^2)$,
which reflects the parameters for the general rule to construct these
two-dimensional lattices \cite{GrunbaumShepard:1987}.
In the present case, every vertex is
surrounded by one square and two octagons. Each lattice site in the
bulk has three neighbours. Along the boundary one of them is replaced by
a virtual neighbour, so that $q_i=3$ throughout.

Non-crossing (nc) diagonal square lattice (\Fref{nocrossing}): This lattice is based
on a square lattice
by adding alternate diagonals such that there are no crossing (nc) diagonals. Each
site has either four or eight nearest neighbours. 
Boundary sites have either one, three or five virtual neighbours.

\begin{figure}
\subfigure[The \nSQUA. $L_x=L_y=6, N=36$.\flabel{square}]{\includegraphics[scale=0.21]{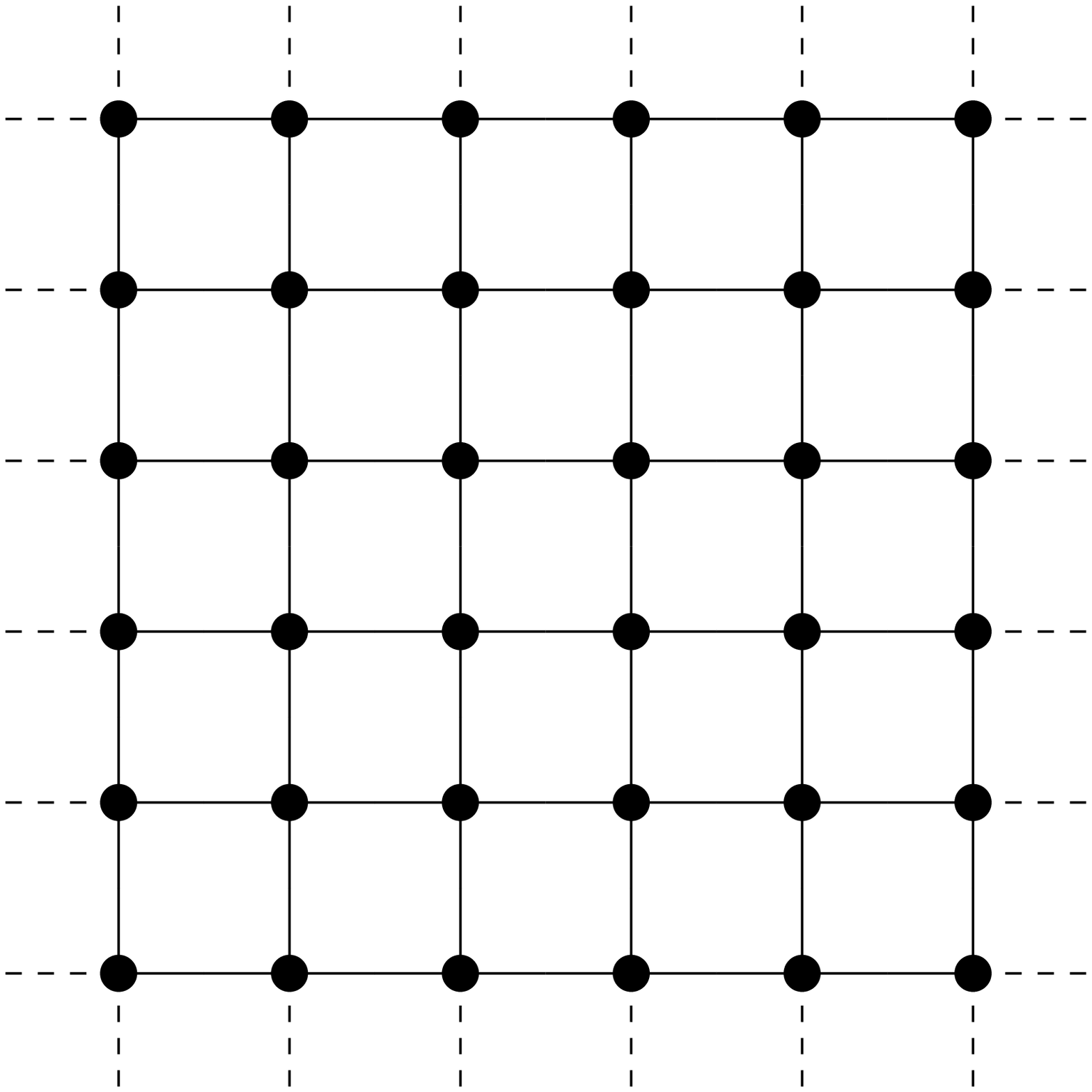}}
\subfigure[The \nJAGG. $L_x=4, L_y=9, N=36$.\flabel{jaggy}]{\includegraphics[scale=0.28]{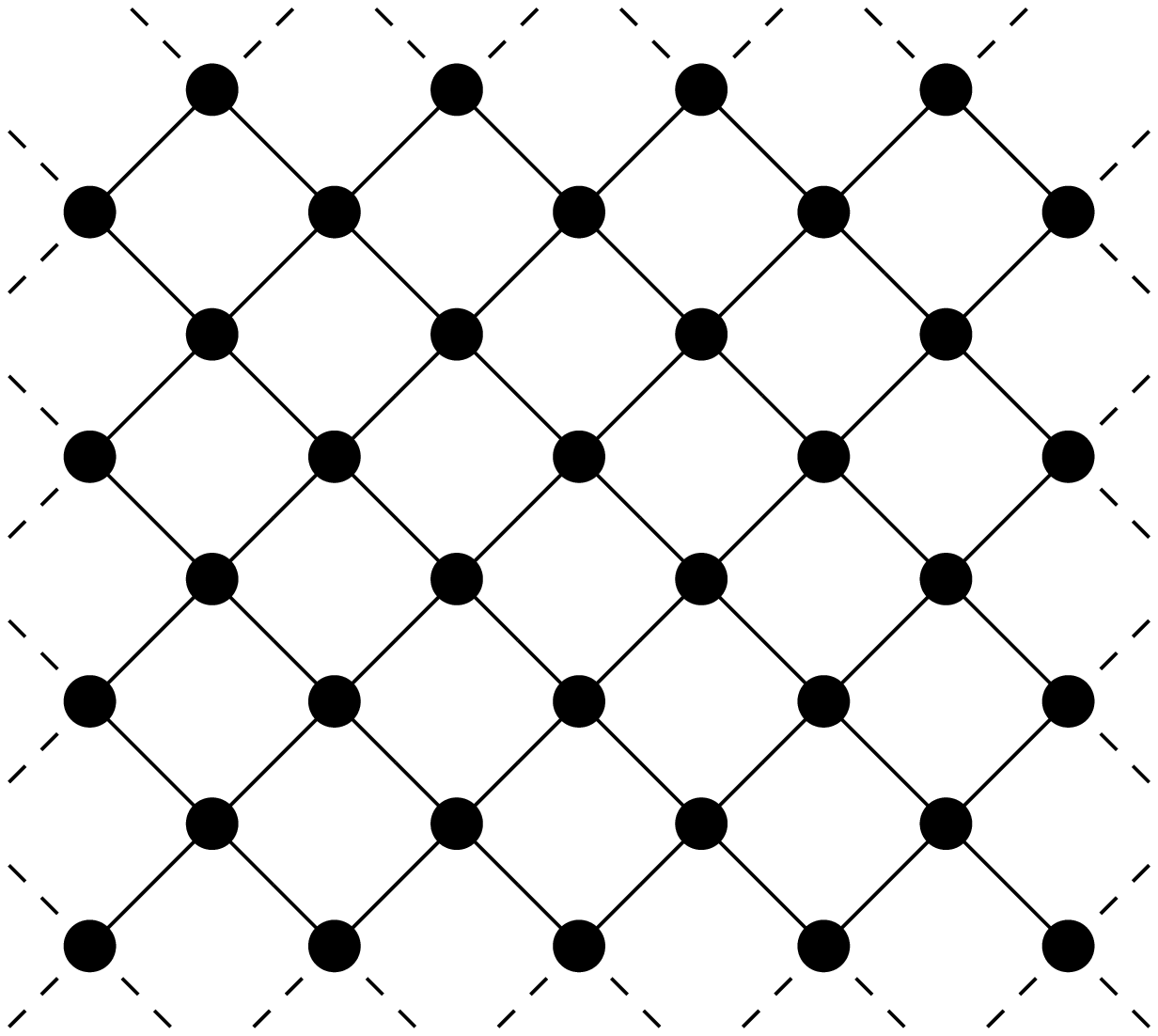}}
\subfigure[The \nARCH. $L_x=8, L_y=4, N=32$.\flabel{archimedes}]{\includegraphics[scale=0.23]{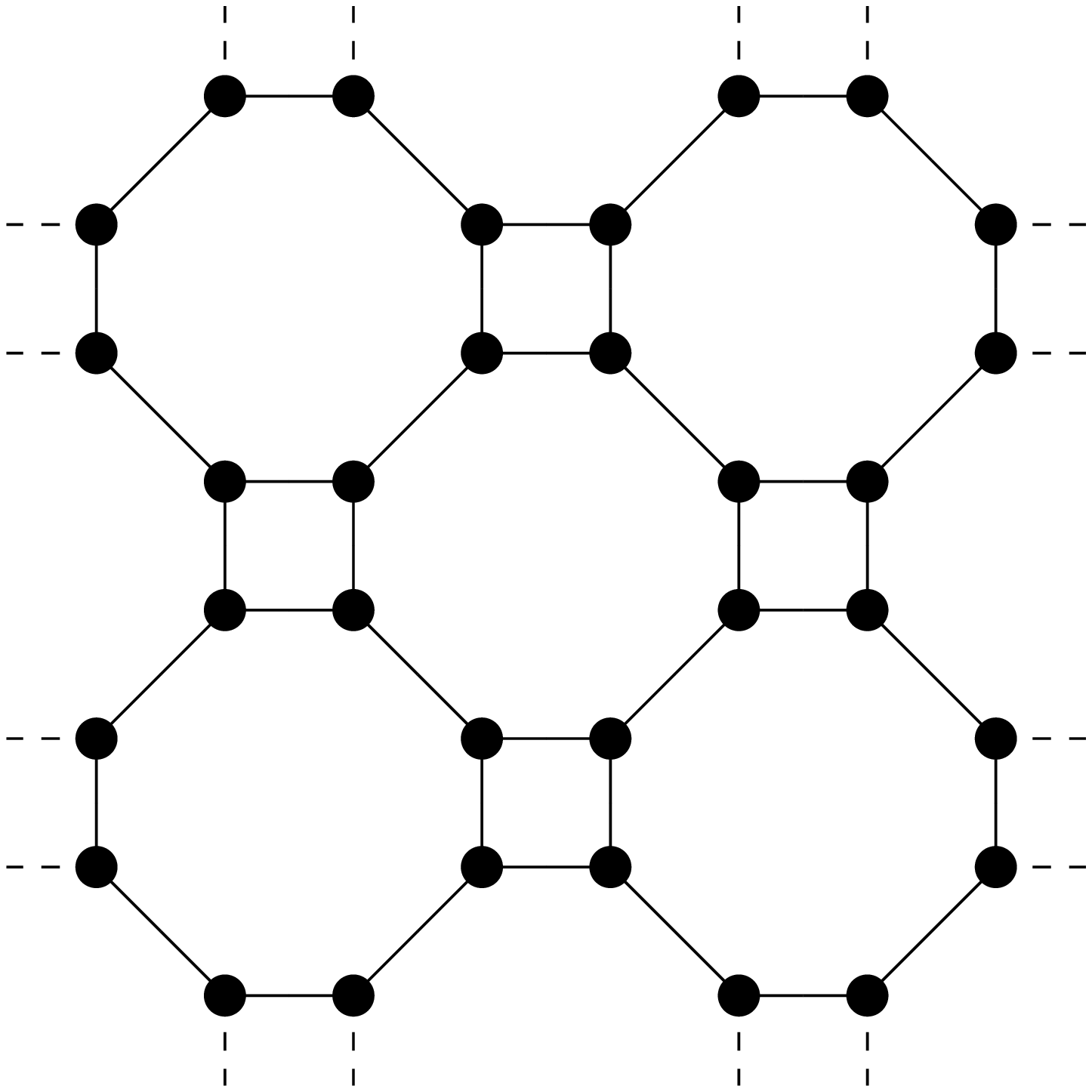}}
\subfigure[The non-crossing (nc) diagonal square\xspace lattice\xspace. $L_x=L_y=5, N=25$.\flabel{nocrossing}]{\includegraphics[scale=0.21]{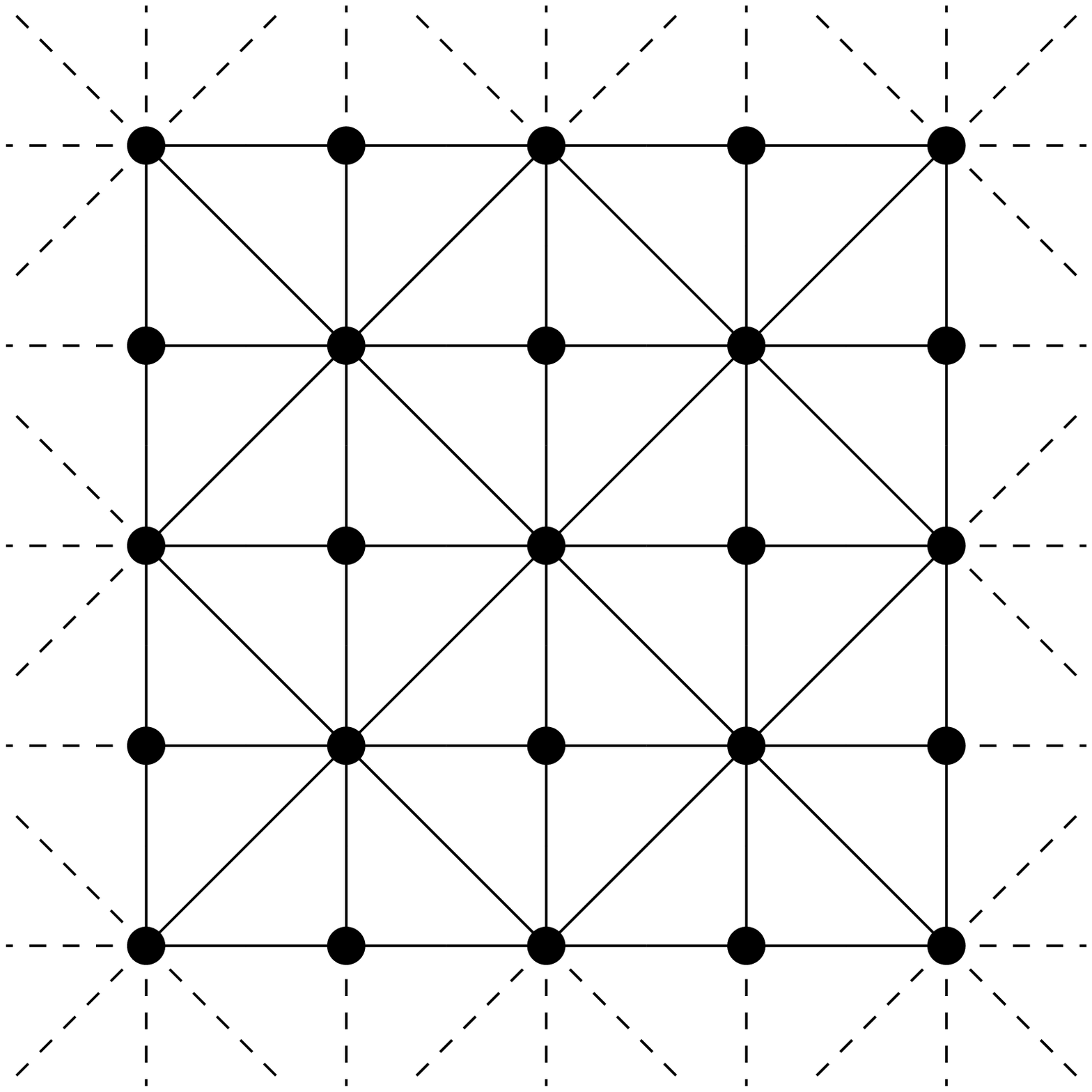}}
\caption{
\flabel{2d_lattices_square}
The four two-dimensional lattices with four-fold symmetry considered in this article.}
\end{figure}

Triangular lattice (\Fref{triangular}): This lattice is the approximate square-shaped
clipping of a tessellation of the plane by triangles. It is probably the second most
frequently studied lattice in statistical mechanics. Owing to its six-fold symmetry
which clashes with the four-fold symmetry of the square, boundary conditions at
the upper and the lower edge differ from those on the left and on the right. This is
not the case for three of the four preceding two-dimensional lattices, but does
similarly apply to the following three. In the present case, the number of virtual
neighbours of sites along the edge varies between one and four, with a constant
$q_i=6$ throughout.

Kagom{\'e} lattice (\Fref{kagome}): This lattice was first studied by Sy{\^o}zi
\cite{Syozi:1951,Syozi:1972}. Its name is Japanese, referring to the pattern of holes (``me'') in a basket (``kago''). Each site in the bulk has four nearest
neighbours.  This lattice has a six-fold symmetry, which generates three
different boundary conditions by the way we decided to cut it at top and
bottom. Similar to the \JAGG, in hindsight we may have picked slightly different
boundaries.

Honeycomb lattice (\Fref{honeycomb}): Similar to the triangular lattice, this lattice
is a tiling by hexagons, leading to a honeycomb-shaped pattern. Each bulk site has
three neighbours, $q_i=3$, and the number of virtual neighbours along the
edge is one everywhere, even when top and bottom edges differ from those on the
left and on the right.

Mitsubishi lattice (\Fref{mitsubishi}): This is Japanese which translates to
``three-diamond'' reflecting the shape of the lattice (the naming is inspired by the
logo of the famous Japanese company of the same name). It is also known as
``the diced lattice'' \cite{Syozi:1951,Syozi:1972}. Each lattice site has either three
or six nearest neighbours with a number of virtual neighbours at the boundary
varying between one and four. Again, top and bottom edges differ from those left
and right.

\begin{figure}
\subfigure[The \nTRIA. $L_x=5,  L_y=7, N=35$.\flabel{triangular}]{\includegraphics[scale=0.20]{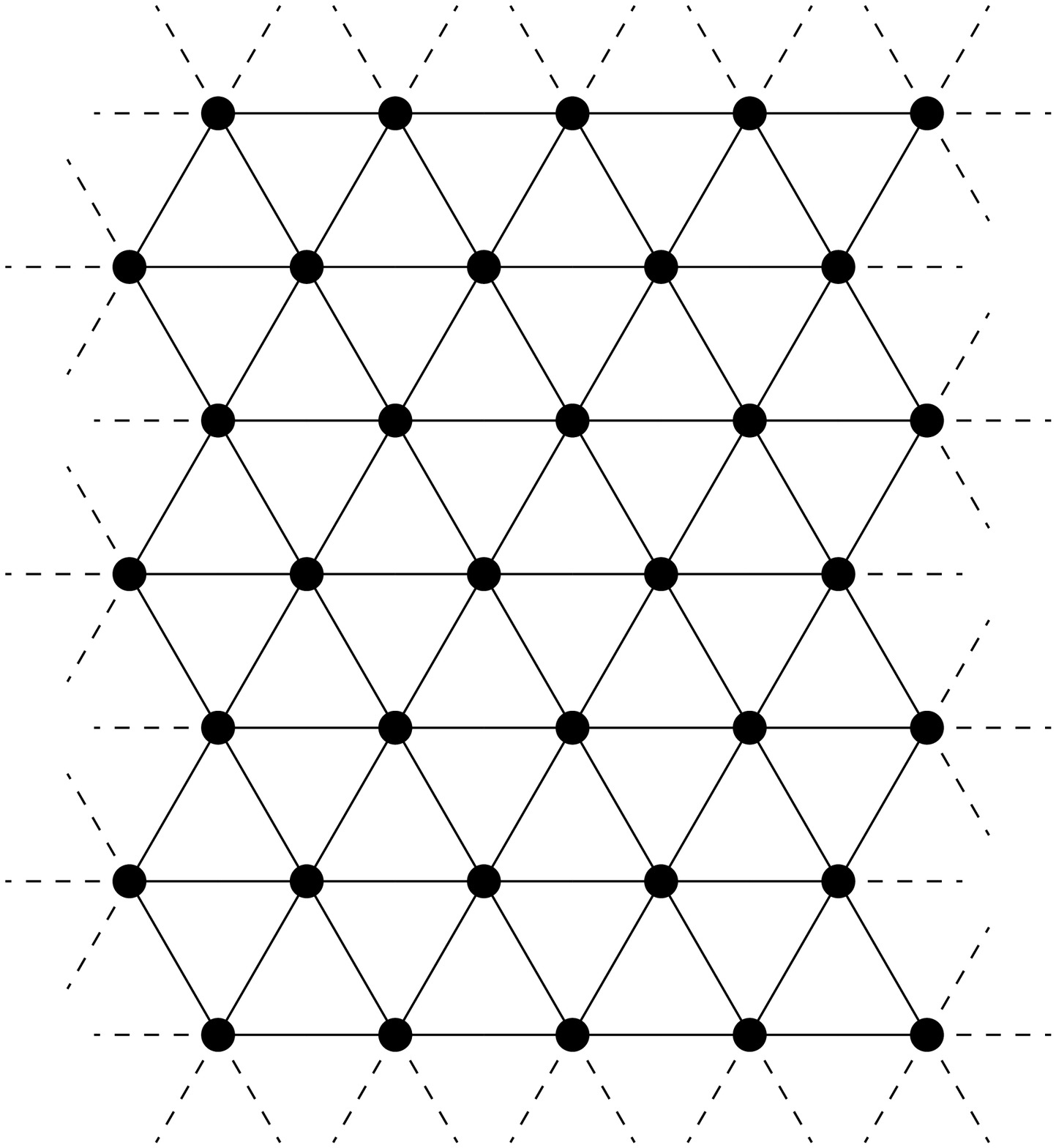}}
\subfigure[The \nKAGO. $L_x=10, L_y=4, N=40$.\flabel{kagome}]{\includegraphics[scale=0.22]{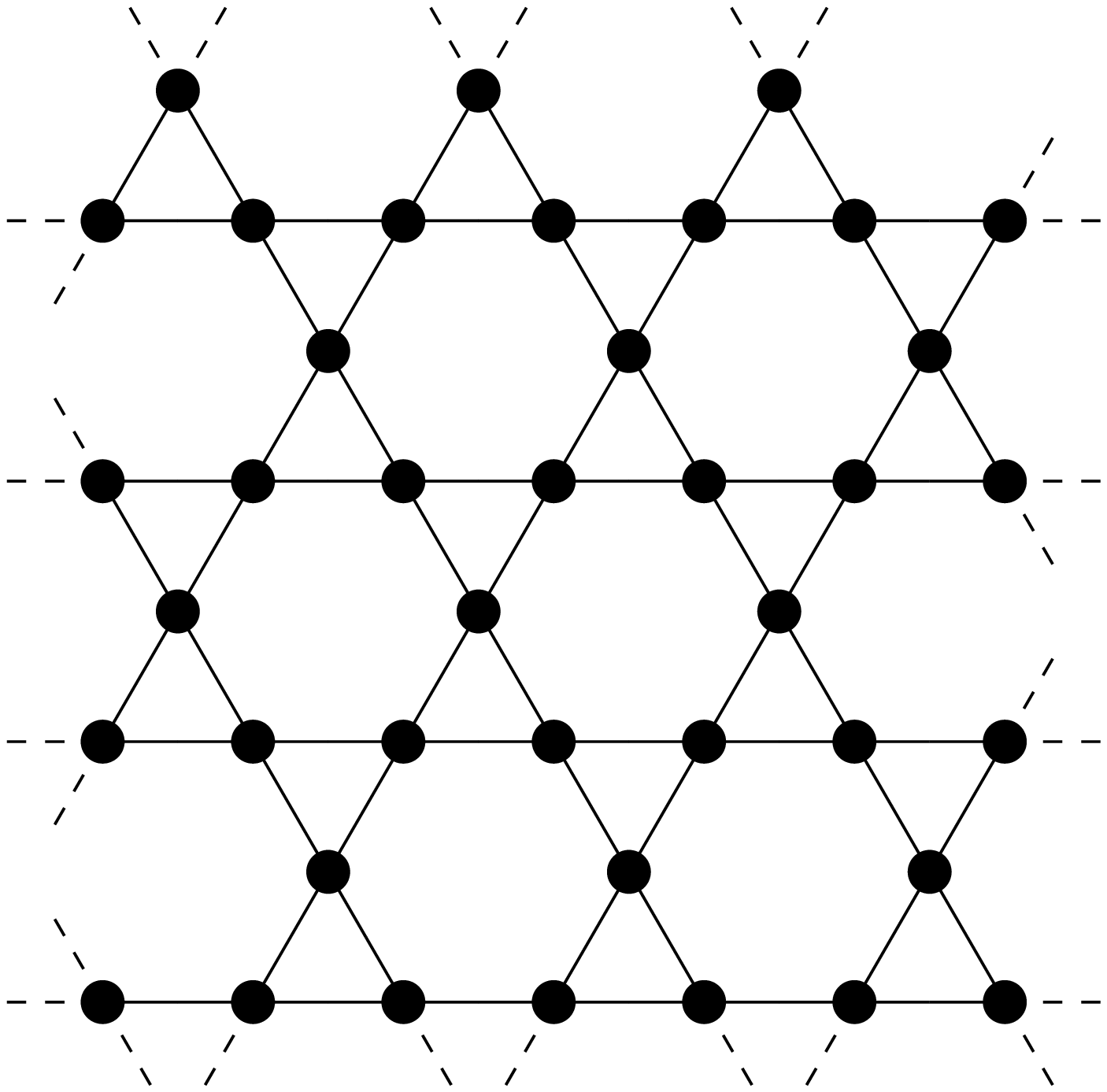}}
\subfigure[The \nHONE. $L_x=9,  L_y=4, N=36$.\flabel{honeycomb}]{\includegraphics[scale=0.22]{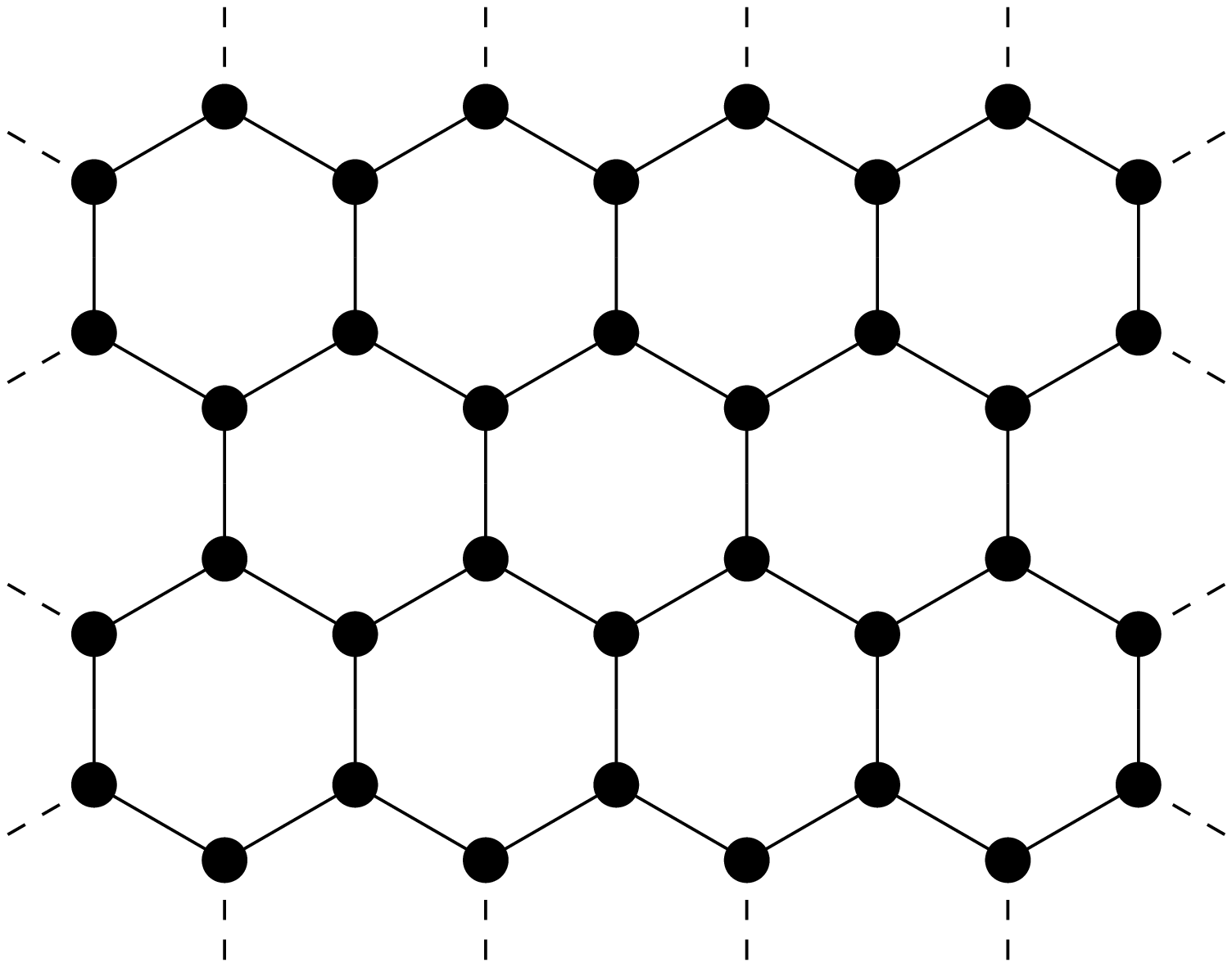}}
\subfigure[The \nMITS. $L_x=5,  L_y=7, N=35$.\flabel{mitsubishi}]{\includegraphics[scale=0.24]{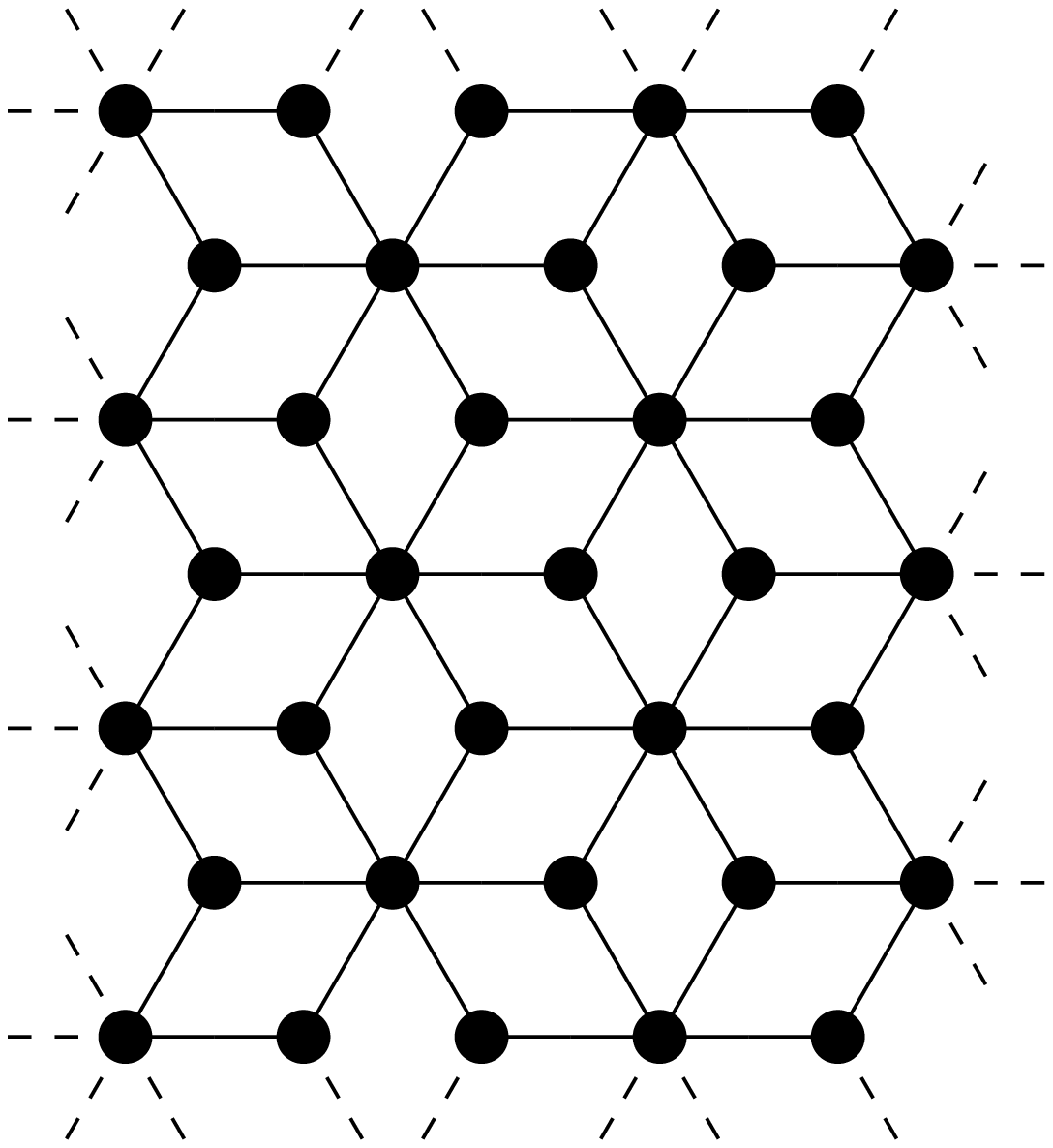}}
\caption{
\flabel{2d_lattices_hex}
The four two-dimensional lattices with six-fold symmetry considered in this article.}
\end{figure}

Some of the two-dimensional lattices are related by transforms, which are
frequently used in the analysis of equilibrium critical phenomena
\cite{Syozi:1951,Syozi:1972,Baxter:2007,Loebl:2010}.  Denoting the
duality transform by $\mathcal{D}$, the star-triangle transform by $\Delta$ and
the decoration-iteration transform by $\mathcal{I}$, the following equalities
hold between two-dimensional lattices:
\begin{eqnarray}
\mathcal{D}(\text{\nHONEshort})&\equiv&\text{\nTRIAshort}\nonumber\\
\mathcal{D}(\text{\nKAGOshort})&\equiv&\text{\nMITSshort}\nonumber\\
\mathcal{D}(\text{\nARCHshort})&\equiv&\text{\nNOCRshort}\nonumber\\
\mathcal{D}(\text{\nSQUAshort})&\equiv&\text{\nSQUAshort}\nonumber\\
\Delta(\text{\nMITSshort})&\equiv&\text{\nTRIAshort}\nonumber\\
\Delta(\mathcal{I}(\text{\nHONEshort}))&\equiv&\text{\nKAGOshort}\nonumber\ .
\end{eqnarray}

\subsection{Implementation details}
Throughout this work we used the same implementation in C of the AMM, which
takes as an input the adjacency information of the various lattices, that are
generated separately from running the actual Manna Model. The full adjacency
matrix is extremely sparse and it therefore makes little sense to store the
information in that format. Rather, all sites are sequentially indexed and sites
adjacent to a given site are listed by their index in a sequence (of varying length).
Negative indices indicate virtual
neighbours. The adjacency information is then filled into a C \verb#struct#
for each site $i$, which contains a vector holding the sequence of
indices of adjacent sites, their number, a flag whether the site has
been hit by the currently running avalanche and finally the height $z_i$. Using
indices rather than pointers to reference sites in our experience
produces very fast code with strongly optimising compilers.

Storing the adjacency information in memory (rather than being implemented
explicitly by rules), makes the code more flexible, but large lattices comparatively
expensive in terms of memory requirements. Significant amounts of memory
memory are also required for the stack of active sites, which holds every site
$i$ whose $z_i$ exceeds $1$. Sites are placed on the stack at the time when they
make the transition from $z_i=1$ to $z_i=2$. Random sequential updating requires
random access to that stack. Sites $i$ picked from the stack topple only once and
thus remain on the stack until $z_i\le1$. At the time when an avalanche is triggered
all sites might hold one particle, so that the theoretical maximum number of sites
exceeding the threshold at any one time simultaneously is $(N+1)/2$. Although this
maximum is not reached in practice, building in safeguards to protect smaller
stacks from overflowing is computationally more costly than providing a stack as large as the theoretical maximum.

A second stack is required to keep track of all sites toppling during any one
avalanche. Once a site is updated, a flag associated with it is changed, and the
site's index is placed on the stack. It is an invariant that all sites with the flag raised
are located on the stack. The area of an avalanche is the height of that stack at the
end of the avalanche. The flags are reset by scanning through the stack.

Even when using memory lavishly, memory requirements for the AMM
implementation as described above are rather modest compared to what any
modern desktop computer has to offer. In one and two dimensions, large lattices are prohibitively
large in terms of CPU time, not in terms of memory, as the average avalanche
size grows quadratically in the linear system size, \ie in the (average) chemical
distance to the open (dissipative) boundary.
An additional,
sometimes very noticeable
constraint on systems with multiple cores or multiple logical cores
(hyper-threading) is the memory bus, which can be alleviated only partly
by reducing the memory requirements.

\subsection{Output}
The output of the code described above is a string of moments,
effectively subsamples, averaged over a number of avalanches (ranging from many millions to
several ten thousand), which we call ``chunks'' in the following. 
Typically, $100$ to $10\,000$ chunks were generated for each lattice.
The
statistics of the chunks allows for an estimate of the statistical
error, while a simple average (weighted by the chunk size if necessary)
across chunks produces an unbiased, consistent estimate of the moments.

The sizes of the chunks were chosen so that a new chunk would be
produced every ten to sixty minutes. In one dimension the linear size of
the lattices spanned about three orders of magnitude, in two dimensions
the square root of that. The average size and thus roughly the CPU time to produce a single avalanche
grows quadratically in the linear size, ranging over six orders of
magnitude in one dimension, and over three in two dimensions, see
\Trefs{system_sizes_1D}{system_sizes_2D}. The chunk sizes have
to be adjusted accordingly. At the same time, the avalanche-avalanche
correlation time increases like a power of the linear extension, but
\emph{not} with the dynamical exponents $z$, which defines the link
between microscopic and macroscopic time scale, but with $L^{D-d}$,
which is a measure of the characteristic fluctuations of the AMM
in the interface picture
\cite{PaczuskiBoettcher:1996,Pruessner_exactTAOM:2003,MorandPruessnerChristensen:2010}.

The chunks were generated on the SCAN facility at Imperial College
London, which harvests CPU time from undergraduate computing facilities
when not used by students, providing up to $706$ logical CPUs
simultaneously, mostly in the form of Intel\textsuperscript{\textregistered}
Core2\textsuperscript{\texttrademark} processors with 2.66 GHz.
In this setup, all I/O is done over the network, which in principle constraints the
amount of output per process running.
Given the drastic
difference in CPU time required to generate a single avalanche, by far
the most CPUs were assigned to the largest systems. In two dimensions,
the smallest four system sizes each had one logical CPU to themselves.
As a result, the smallest systems (that we kept, see below) 
have sample sizes exceeding $10^{9}$ avalanches,
which leads to highly accurate estimates for their momenta.
In total, about $168\,415$ and $72\,451$ hours of CPU time were spent on
generating the data in one and two dimension, respectively.

For simplicity, checkpointing was implemented only right after a chunk
was written out. The power cycling of the computer setup thus limited
the amount of time available to generate a single chunk.
While the original intention was to choose the chunk size such that
correlations (which normally come in the form of anti-correlations) are
negligibly small, this turned out to be unsustainable for the very large
system sizes. However, uncorrelated chunks greatly facilitate the
calculation of statistical errors, compared to, say, a full-blown
resampling plan \cite{Efron:1982}. To this end, chunks were merged
during post-processing, as discussed below.

As each instance of the AMM was started with an empty lattice, a generous amount
of chunks was dropped as transient from the set considered in the subsequent data
analysis. This equilibration ``time'' was estimated by inspection of
individual series of chunks, as a multiple of the time to ``obvious''
stationary. 
For the largest lattices in one dimension typically $10^{5}$ avalanches 
(more for larger lattices) were rejected as 
transient and $10^{6}$ retained for statistics in one dimension. In two 
dimensions, typical numbers were
$6\cdot 10^{6}$ as transient and $400 \cdot 10^{6}$ for statistics\footnote{In
hindsight, the lattices used in two dimensions appear rather small compared to
one, as the CPU time required is essentially determined by the linear extent,
$N^{1/2}$.}.
As a rule of thumb, at least $3/2$ times the number of sites in the lattice
avalanches were removed as transient. For smaller lattices a much larger
fraction was taken, as for them, equilibration was often apparently 
reached within a single chunk. Equilibration can also be observed in the 
density of particles, which, at least in one dimension, initially grows almost perfectly 
linearly with only a minute amount of dissipation. Once a certain fraction of
sites is occupied, the occupation density displays very litte relative fluctuations,
for large lattices of around $2\cdot10^{-4}$.

The transient serves the additional purpose of warming up the random
number generator (RNG), which was of course seeded uniquely in each
instance, except for the \nMITS, whose
190 different seeds equal those used for \nHONE. All results presented in the following are based on
the Mersenne twister \cite{MatsumotoNishimura:1998a}, which has received
some criticism
for its correlations across differently seeded instances
\cite{Marsaglia:2005}. The independence of the present results from the
RNG
was tested by re-running a few setups with Marsaglia's KISS RNG
\cite{Marsaglia:1999}.

\subsection{Statistical error}
All results presented below are based on moments of the avalanche size,
area and duration. 

The first moments of all three observables generally posed a problem,
possibly
because they were determined with an accuracy so high that it was 
virtually impossible to account for their corrections to scaling
(see below). The statistical error of all moments was
calculated on the basis of chunks whose size was chosen as to ensure their independence.
This was tested by firstly calculating their autocorrelation
function and, secondly, by successively increasing their size, \ie
merging them, to probe whether the statistical error 
derived on their basis
was affected by
this operation. In fact, for the largest system sizes (and thus smallest
chunk sizes) tested, some 
correlations were visible (correlation length of about $0.7$ chunks), and we decided to merge
ten consecutive chunks throughout. It turned out, however, that only the
statistical errors of the first moments were noticeably, yet still
insignificantly, affected at all by this operation.

It is straight-forward to determine estimates and their estimated
variance on the basis of independent chunks. Denoting the observable in
the $i$th chunk by $c_i$, an unbiased, consistent estimator \cite{Brandt:1998}
of its
population average $\ave{c}$ is
\begin{equation}
\frac{1}{M_c}\sum_{i=1}^{M_c} c_i
\end{equation}
given a sample size of $M_c$. In the following, we will denote that
estimate itself by $\ave{c}$. An unbiased, consistent estimator of the
variance of this estimator is
\begin{equation}
\frac{1}{M_c-1} \left[
\frac{1}{M_c} \sum_{i=1}^{M_c} c_i^2
-
\left( \frac{1}{M_c}\sum_{i=1}^{M_c} c_i\right)^2
\right] \ .
\end{equation}
These estimates were used as input values for the estimation of the
exponents, based on the scaling assumption discussed in the following.
For completeness, covariances of the estimators for the averages of 
two observables $c$ and $c'$, as used in \Eref{err_gn}, were estimated using
\begin{multline}
\cov{c,c'}=\\
\frac{1}{M_c-1} \left[
\frac{1}{M_c} \sum_{i=1}^{M_c} c_i c_i'
-
\left( \frac{1}{M_c}\sum_{i=1}^{M_c} c_i\right)
\left( \frac{1}{M_c}\sum_{i=1}^{M_c} c_i'\right)
\right] \ .
\elabel{covar}
\end{multline}
As detailed below, 
the finite size scaling of the moments of each observable was determined
in the form
\begin{equation}
\ave{x^n} = a_{x,n} L^{\fssExpo{x}{n}} + \textrm{corrections} \ 
\elabel{fss_leading_order}
\end{equation}
where $x$ stands for the size $s$, the area $a$ or the duration $t$.
The combined information of the scaling of a number of different moments
$n$ was then used to estimate the finite size scaling exponents. 
Even when we implemented the first eight moments, we
used only moments $2$ to $4$ 
(\eg
$\fssExpo{s}{2}$,
$\fssExpo{s}{3}$,
$\fssExpo{s}{4}$)
when fitting the avalanche dimension
(using $2=(2-\tau)D$ to determine $\tau$) and moments $2$ to
$5$ to fit exponents characterising the scaling of avalanche area and
duration (all exponents are defined below).
Different moments estimated on the basis of a single Monte-Carlo simulation are not independent. To account for that, we (rather generously) multiplied the
statistical error of each moment by the square root of the number of
moments considered simultaneously, as if each moment was determined
independently from the others \cite{MoloneyPruessner:2003b}. For example, the statistical error of
the second, third and fourth moment of the avalanche size was multiplied
by $\sqrt{3}$, before $\fssExpo{s}{2}$, $\fssExpo{s}{3}$ and
$\fssExpo{s}{4}$ respectively were determined. It seems that these
correlations are frequently ignored in the literature. This
procedure does \emph{not} account for correlations in the estimates of
moments of different observables and thus our results for the finite
size scaling exponents for different observables are not independent.
Multiplying again by the square root of the
number of different observables considered, may, however, seriously
overestimate the impact of these correlations.

\subsection{Finite size scaling}
Making the usual finite size scaling assumption for observable $x$, its
probability density function $\pdf{x}{x}$ follows
\begin{equation}
\pdf{x}{x;L} = a_x x^{-\tau_x} \GC_x\left(\frac{x}{b_x L^{D_x}}\right)
\elabel{power_law_distribution}
\end{equation}
asymptotically in large $x\gg x_0$ with lower cutoff $x_0$, linear system size $L$,
non-universal metric factors $a_x$ and $b_x$ and universal exponents $\tau_x$ and
$D_x$. Below, the fractal dimension of avalanches is denoted by $D_a$
\cite{Luebeck:2000} and the avalanche area exponent by  $\tau_a$.
For historic reasons, 
the dynamical exponent is denoted by $z$ (rather than $D_t$), the avalanche
duration exponent by $\alpha$ (rather than $\tau_t$), and finally the 
avalanche dimension by $D$ (rather than $D_s$) and the avalanche size
exponent by $\tau$ (rather than $\tau_s$).
The universal, dimensionless scaling function
$\GC_x$ of a dimensionless argument
decays, for large arguments, faster than any power law, so that all moments
\begin{equation}
\ave{x^n}(L) = \int_0^\infty \dint{x} \pdf{x}{x;L} x^n
\elabel{moment_definition}
\end{equation}
exist for any finite system and $n\ge0$. Provided that $n+1-\tau>0$ one can easily show 
\cite{DeMenechStellaTebaldi:1998,ChristensenETAL:2008}
that gap scaling follows, so that 
\begin{equation}
\ave{x^n}(L) = a_x \left(b_x L^{D_x}\right)^{n + 1 - \tau_x}
\int_0^\infty \dint{y} y^{n-\tau_x} \GC_x(y)
\elabel{xn_asymptote}
\end{equation}
asymptotically in large $L$, or, more accurately,
\begin{equation}
0< \lim_{L\to\infty} \frac{\ave{x^n}(L)}{L^{D_x (n + 1 - \tau_x)}} <
\infty \ .
\elabel{proper_def_of_expo}
\end{equation}
To determine $D_x$ (and $\tau_x$, if independent), the
leading order scaling of the moments according to
\eref{fss_leading_order} was estimated,
fitting the resulting exponents $\fssExpo{x}{n}$ against $D_x (n + 1 -
\tau_x)$, without allowing for any further corrections. The results are
shown in \Trefs{avalanche_exponents}{Exponent_summary}.

Because of particle conservation in the bulk, every
particle placed on the lattice anywhere can only leave through
the boundary. On the way there, it performs an independent random walk, 
even when it occasionally rests \cite{NakanishiSneppen:1997}.
Every move by a particle is caused by a site's toppling and the number
of particle moves during an avalanche is therefore exactly twice the
number of topplings.
At stationarity one particle leaves the system per particle added and
(half) its trajectory length is its contribution to the various avalanches it
has been part of. The average contribution per particle and thus per
avalanche is the average trajectory length, which scales like $L^2$
independent of the dimensionality of the lattice \cite{ItzyksonDrouffe:1997}
and many of the details of the boundary. 
This argument remains valid if avalanches of size $0$ are excluded 
from the average (which 
we did \emph{not}), provided the probability of producing an avalanche of 
size $0$ (\ie hitting an empty site) does not converge to $1$.

As a result $\ave{s}\propto L^2$ asymptotically, or, here, $\ave{s}\propto N^{2/d}$, \ie $\fssExpo{s}{1}=2$ and under the
assumption of gap scaling $2=(2-\tau)D$. This identity has been
used in the fitting of the avalanche dimension, \ie $\fssExpo{s}{n}$ for
$n=2,3,4$ was fitted against $2+D(n-1)$. At the same time, comparing the
estimate for $\fssExpo{s}{1}$
(\Trefs{avalanche_exponents}{Exponent_summary})
to the exact value $2$ allowed us to
assess the fitting procedure, in particular the form of the corrections
discussed below.

Defining
\begin{equation}
-\Sigma_x=D_x(\tau_x-1)
\elabel{def_Sigma}
\end{equation}
the assumption of a sufficiently
\cite{ChessaETAL:1999,JensenPruessner:2004} narrow joint probability
density function
\cite{JensenChristensenFogedby:1989,ChristensenFogedbyJensen:1991,Luebeck:2000}
produces the scaling law $\Sigma:=\Sigma_s=\Sigma_t=\Sigma_a$, \ie
\begin{equation}
-\Sigma=D(\tau-1)=z(\alpha-1)=D_a(\tau_a-1)\ .
\elabel{narrow_joint}
\end{equation}
As this is not a mathematical identity \cite{PaczuskiBoettcher:1996} (in
fact, it seems to be broken in the original Manna Model \cite{Luebeck:2000}) it
is tested as a scaling hypothesis below. The exponent $\Sigma$ can be seen 
as a replacement for the exponents $\tau$, $\alpha$ and $\tau_a$, as
$\fssExpo{s}{n}=nD+\Sigma$, 
$\fssExpo{t}{n}=nz+\Sigma$, 
$\fssExpo{a}{n}=nD_a+\Sigma$ for $n>\tau-1$, $n>\alpha-1$ and $n>\tau_a-1$ respectively.

Finally, as in the BTW model \cite{ChristensenOlami:1993}, 
in the Manna Model avalanches are often assumed to be compact
\cite{Ben-HurBiham:1996,ChessaETAL:1999}, \ie $D_a=d$ (supposedly up to
the upper critical dimension, where we expect dangerous irrelevant
fields to spoil this scaling relation).
There is no mathematical proof for this feature, yet numerically it is well
verified, see \Trefs{avalanche_exponents}{Exponent_summary}. Clearly $D_a\le d$, 
which served as another constraint to
assess the quality of the exponents extracted.

\subsubsection{Corrections to scaling}
The exponents $\fssExpo{x}{n}$ characterise the asymptotic scaling of
the moments in large $L$. It is widely known \cite{Wegner:1972}, however, 
finite lattices suffer from finite size corrections, which manifest
themselves as sub-leading terms to be included on the right hand side of
\Eref{fss_leading_order} \cite{ChessaETAL:1999}. A priori, the structure of the corrections is
not known, yet they have a marked impact on the quality of the results,
as they are \emph{imposed} when fitting the data.

Given that there is often no natural way of defining the linear extent
of a lattice, we decided to replace $L$ (as in \Eref{fss_leading_order}) 
by $N^{1/d}$.
We considered a host of different fitting functions, such as
\begin{eqnarray}
\ave{x^n}(N) &=& A_{x,n} N^{\fssExpo{x}{n}/d} \elabel{fit_No}\\
\ave{x^n}(N) &=& A_{x,n} N^{\fssExpo{x}{n}/d} + B_{x,n} N^{\fssExpo{x}{n}/d-\omega_{x,n}/d} \elabel{fit_D}\\
\ave{x^n}(N) &=& A_{x,n} N^{\fssExpo{x}{n}/d} + B_{x,n} N^{\fssExpo{x}{n}/d-1/4} \elabel{fit_OneQuarter}\\
\ave{x^n}(N) &=& A_{x,n} N^{\fssExpo{x}{n}/d} + B_{x,n} N^{\fssExpo{x}{n}/d-1/2} \\
&& \hspace{2.2cm} + C_{x,n} N^{\fssExpo{x}{n}/d-1}  \nonumber
\end{eqnarray}
and eventually settled for
\begin{multline}
\ave{x^n}(N) = A_{x,n} N^{\fssExpo{x}{n}/d} 
+ B_{x,n} N^{\fssExpo{x}{n}/d-1/4}\\
+ C_{x,n} N^{\fssExpo{x}{n}/d-1/2}
\elabel{fit_TwoQuarter}
\end{multline}
which yielded particularly good estimates. In particular, $\fssExpo{s}{1}=2$ was 
reproduced quite reliably. The quality of the estimates
was assessed on the basis of the goodness of fit determined in the
Levenberg-Marquardt least
square fitting routine \cite{PressETAL:1992}. For the vast majority of moments and lattices,
we could have dropped the last term in \Eref{fit_TwoQuarter} arriving \Eref{fit_OneQuarter} and still achieved a goodness of fit (q-value) 
of more than $0.9$. However, the first moments of the avalanche size, whose
finite size scaling exponent
is the only exactly known one, was particularly poorly fitted without that term. For
consistency, we decided to fit \emph{all} moments using \Eref{fit_TwoQuarter}, achieving
typically $q$-values of better than $0.9$, suggesting that we overestimated the
statistical errors. In all result tables, fits that had a $q$-value of less than $0.1$ are marked as
such.

To reduce the impact of a possible dependence on the (arbitrary) choice of the initial values of
the free parameters, the number of terms in the fitting function was
increased successively starting from \Eref{fit_No}, using the
estimates of the parameters in the previous fit as the initial value of
the same parameters in the next.

The simplicity of \Eref{fit_TwoQuarter} meant that we had to drop results for small
system sizes, which suffer from stronger finite size corrections, yet
were determined with much greater accuracy than those of the bigger systems. This
is a common theme in the present work: Moments in small system sizes
were determined with such great accuracy that very many (a priori unknown) correction terms
would have to be included to account for all such details. At the same time,
it makes little sense to have almost as many free parameters in the
fitting function as there are data points to fit. In order to retain
the goodness of fit as a meaningful device to determine the quality of
the fit, we therefore removed the smallest four system sizes from the
procedure (in one and two dimensions), keeping the seven system sizes
listed in \Trefs{system_sizes_1D}{system_sizes_2D}.

Increasing the system size in order to suppress correction terms comes
at the price of increased relative error if $\tau_x>1$. According to
\Eref{fss_leading_order} and \eref{proper_def_of_expo}, the variance
of the $n$th moment has leading order $L^{D_x(2n+1-\tau_x)}$, \ie the
relative error scales like $L^{D_x(\tau_x-1)/2}$. Moreover, correlations
are expected to die off after $L^{D-d}$, which reduces the number
of effectively independent measurements with increasing system size.

\subsection{Moment ratios}
Ratios of products of moments, which (to leading order) are independent from $L$,
characterise the scaling function $\GC_x$,
\Eref{power_law_distribution}, directly. In equilibrium phase transition, the
so-called Binder-cumulant \cite{Binder:1981a,Binder:1981b} is the best known such ratio, signalling the
deviation from a Gaussian distribution of the order parameter around the
critical point. There are many ways of constructing suitable moment
ratios; assuming $\ave{x^n}\propto L^{D_x (n + 1 - \tau_x)}$, it is easy
to see that any ratio of products of moments, which has the same number
of moments (to cancel $\tau_x\ne1$) and the same sum of orders of moments
(to cancel $D_x$) in numerator and
denominator leads to a non-scaling quantity. Since the second moment is
positive and bounded away from $0$, traditionally moment ratios are
formed by dividing by a power of it. Moreover, in many phase
transitions, the order parameter follows a distribution with $\tau_x=1$,
which removes the constraint of having the same number of moments in the
numerator and the denominator.

While the sets $\ave{x^{n-m}}\ave{x^{n+m}}/\ave{x^{n}}^2$ attracts by its
simplicity and symmetry, the set
\begin{equation}
\momrat{x}{n} = \frac{\ave{x^n}\ave{x}^{n-2}}{\ave{x^2}^{n-1}} \ ,
\elabel{moment_ratio}
\end{equation}
has the particularly nice feature that $\momrat{x}{1}=\momrat{x}{2}=1$ by
definition, which fixes the metric factors $a_x$ and $b_x$ in
\Eref{power_law_distribution} by \emph{imposing} for $n=1,2$
\begin{equation}
\momrat{x}{n} = \int_0^{\infty} \dint{y} y^{n-\tau_x} \GC_x(y) \ ,
\end{equation}
see \Eref{xn_asymptote}
which is then consistent with \Eref{moment_ratio} for all $n$.

The statistical error $\sigma\left(\momrat{x}{n}\right)$ of the estimator of \Eref{moment_ratio} is to
leading order in the sample size
given by
\begin{widetext}
\begin{multline}
\sigma^2\left(\momrat{x}{n}\right) = 
(\momrat{x}{n})^2
\Bigg(
          \frac{\covx{n}{n}}{\left(\ave{x^n}\right)^2}
+ (n-2)^2 \frac{\covx{1}{1}}{\left(\ave{x}\right)^2}
+ (n-1)^2 \frac{\covx{2}{2}}{\left(\ave{x^2}\right)^2} \\
- 2(n-2)(n-1) \frac{\covx{2}{1}}{\ave{x}\, \ave{x^2}}
- 2(n-1)      \frac{\covx{2}{n}}{\ave{x^2}\, \ave{x^n}}
+ 2(n-2)      \frac{\covx{1}{n}}{\ave{x}\, \ave{x^n}} 
\Bigg) \ ,
\elabel{err_gn}
\end{multline}
\end{widetext}
which matches perfectly (typically the first three or four significant digits) the error
as found by the subsampling using chunks, \ie determining $\momrat{x}{n}$ for
each chunk and estimating the error by the square root of its variance over the
number of chunks. In \Eref{err_gn} $\ave{x^n}$ strictly denotes the
\emph{estimator} of the $n$th moment  and $\cov{x^n,x^m}$ the estimated
covariance of the $n$th and $m$th moment, see \Eref{covar}. Consistent with the
preceding discussion, we used averages and statistical errors derived from chunks.

All lattices were set up with the intention of creating an aspect ratio of $1$,
which is trivial as long as the lattice has a four-fold symmetry. In particular for
lattices without that symmetry, such as
the \TRIA, the \KAGO, the \HONE and the \MITS, but also, say, the \JAGG, the aspect ratio might 
deviate slightly from unity and converge to $1$ only with increasing system size.
In any case, the aspect ratio might
be more reasonably be
defined using the Manhattan distance across the lattice. It is well
known that universal scaling exponents are
generally independent from the aspect ratio, whereas finite size scaling
functions are not \cite{PrivmanHohenbergAharony:1991}.
Therefore, deviations of the moment ratios in particular in case of the lattices 
listed above are expected (but did in fact not materialise). 
Surprisingly, even when there is every reason to assume that no such problem can
occur in one dimension, their moment ratios proved particularly 
difficult to analyse.

\section{Results}
\slabel{results}
\subsection{Avalanche exponents}
After stripping off the transient, merging chunks as discussed above,
deriving average moments and errors using the procedures described
above, the scaling exponents $\fssExpo{x}{n}$ where fitted using
\Eref{fit_TwoQuarter} for the three different
observables, $x=s,t,a$, avalanche size, duration and area respectively.
As mentioned above, correlations between moments were taken into account
by multiplying the statistical error of the moment by the square root of
the number of moments considered.
Allowing for no further corrections, $\fssExpo{s}{n}$ were fitted
for each of the twelve different lattices separately
against 
\begin{equation}
\fssExpo{s}{n}=2+D(n-1)
\elabel{fitting_s}
\end{equation}
for $n=2,3,4$. The results are collected in \Tref{avalanche_exponents}. The scaling law $2=(2-\tau)D$ used in \Eref{fitting_s}
was probed independently; $\fssExpo{s}{1}=2$ is a mathematical
identity, but valid only \emph{asymptotically} and the deviation of
$\fssExpo{s}{1}$ from $2$ can therefore serve as an indicator to assess
the quality of the fitting routines and the data and can help confirming
that ``asymptotia is reached''. 
The estimate for $\fssExpo{s}{1}$ on the basis of \Eref{fit_TwoQuarter} is shown
in \Tref{avalanche_exponents} alongside the other finite size scaling exponents.
The exponent $\tau$ stated in 
\Tref{avalanche_exponents} is derived from the estimate of $D$ using
$\tau=2-2/D$.

For all observables (size, area and duration), the first moments turned
out to be problematic. According to \Eref{moment_definition}
moments $n<\tau-1$ remain finite in the
thermodynamic limit, consistent with our observation that smaller
moments generally require more correction terms. The average avalanche
size is particularly difficult to handle, which is determined, due to the presence of
anti-correlations
\cite{Pruessner_exactTAOM:2003,MorandPruessnerChristensen:2010}, with
incredible precision, typically with a relative error of the order $10^{-5}$.
We therefore decided to omit first moments from the determination of the finite
size scaling exponents throughout.

For $x=t$ and $x=a$ no scaling laws were used (even
when $D_a=d$ is generally assumed to hold) and so the exponents $\fssExpo{x}{n}$ were
fitted against:
\begin{eqnarray}
\fssExpo{t}{n}&=&z(n+1-\alpha)\\
\fssExpo{a}{n}&=&D_a(n+1-\tau_a)
\end{eqnarray}
for $n=2,3,4,5$. 

Fitting moments beyond $n=5$ proved very difficult. We decided to drop all
moments beyond the fourth for avalanche size and beyond the fifth for
the avalanche area and duration, as $\fssExpo{x}{n}$ became clearly
dependent on the choice of the initial values of the fitting parameters.
As mentioned above,
in all fitting schemes used, we increased the number of free parameters
successively and used the estimates of the fitting parameters of one 
scheme as the initial values for its extension. For example, 
we used $A_{x,n}$ and $\fssExpo{x}{n}/d$ from a fit against \Eref{fit_No} as 
initial values in a fit against \Eref{fit_OneQuarter}, which in 
turn produced the initial values of $A_{x,n}$, $B_{x,n}$ and 
$\fssExpo{x}{n}/d$ to fit with \Eref{fit_TwoQuarter}. We observed this procedure in
all fitting schemes discussed below.

As the variance of the $n$th moment scales like $L^{D_x(2n+1-\tau_x)}$,
its numerical estimate is increasingly affected by the floating point
precision (double-extended throughout) --- equivalently, the typical largest
measurement of the $n$th moment scales like the $n$th power of the cutoff,
$L^{n D_x}$, which for 
$n=6$, $D_x=2.25$ and $L=2^{16}$
is $2^{216}$. Given that the smallest event size is $0$, 
this is
to be compared to the $64$ bits in the mantissa on a \texttt{long
double} on the x86 architecture.

The exponent $\Sigma_x$ can be derived either from the definition \Eref{def_Sigma}
through $D_x$ and $\tau_x$
or by independently fitting $\fssExpo{x}{n}$ against $nD_x+\Sigma_x$. 
We took that approach for $x=t$ and $x=a$.
Using the three different
observables size, duration and area, provides effectively estimates for
$D(\tau-1)$, $z(\alpha-1)$ and $D_a(\tau_a-1)$ respectively. In the case
of $D(\tau-1)$ this is in fact exactly $D-2$, since we imposed $D(2-\tau)=2$ 
when estimating $D$. The entry for $\Sigma_s$ in \Tref{avalanche_exponents} 
is therefore derived from the estimate for $D$. Except for $\tau$, all other
entries in \Tref{avalanche_exponents} are based on fitting $\fssExpo{x}{n}$ directly.

\Tref{avalanche_exponents} provides very strong support for universality across 
different lattices. Under the assumption that universality holds, 
estimates for exponents gained from different lattices
can be taken together to produce an overall estimate. 
The result of that procedure is shown in \Tref{Exponent_summary}.
As discussed further below, 
we regard only the data referring to the fitting function \Eref{fit_TwoQuarter}, shown
in bold, as reliable, producing the most consistent and robust results.

Fits with \Eref{fit_TwoQuarter} produced very large $q$-factors, suggesting we had been too generous
either with estimating the statistical errors or with the number of free parameters.
We therefore also tried \Eref{fit_OneQuarter},
which, however, gave partly inconsistent results. In particular estimates for
$\fssExpo{s}{1}$ deviated from the exact value $2$ by about $30$ standard
deviations (although the relative error was only $3\cdot10^{-3}$). For the \nLINE
the avalanche size moments tested displayed a poor quality of fit using \Eref{fit_OneQuarter}, as did 
the first moments of the avalanche size for all lattices except for the \NNN, the \FUTA 
and the \ARCH. The results for the fits against \Eref{fit_OneQuarter} are also
summarised in \Tref{Exponent_summary}. A bracket $\qoflow{\cdot}$ indicates
finite size scaling exponents not being fitted with a goodness of fit
better than $0.1$.

In general, the \nFUTA and the \nLINE 
were particularly difficult to fit using whichever fitting function.

In order to extract the exponent of the sub-leading terms, the remainder
$\ave{x^n}(N) - A_{x,n} N^{\fssExpo{x}{n}/d}$ was fitted against
\begin{equation}
\Btilde_{x,n} N^{\fssExpo{x}{n}/d-\omega_{x,n}/d}
\end{equation}
determining $\omega_{x,n}$. This procedure was aiming much more at a qualitative result rather than
a quantitative one and generated rather noisy estimates.
The \nFUTAshort and the \nTRIAshort lattices proved particularly difficult to handle.
The data in $d=1$ produced, unfortunately quite inconsistently
$\omega_{s,n}\approx 0.28$, $\omega_{t,n}\approx0.20$ and $\omega_{a,n}\approx0.20$,
while $d=2$ produced, slightly more consistently 
$\omega_{s,n}\approx 0.23$, $\omega_{t,n}\approx0.32$ and $\omega_{a,n}\approx0.47$ 
fairly independent of lattice and $n$ (but more reliably for large $n$ and observables other than the avalanche size). 
These exponents could in turn be used in \Eref{fit_D} to fit 
the data for $\fssExpo{x}{n}$ at fixed $\omega_{x,n}$. The resulting overall estimates for the
finite size scaling exponents are 
also shown in \Tref{Exponent_summary}.

\begin{table*}[t]
\caption{
\tlabel{avalanche_exponents}
Summary of all exponents characterising the avalanching in the twelve different lattices,
using \Eref{fit_TwoQuarter}.
The estimates for $\tau$ and $D(\tau-1)$ were \emph{not} determined
by fitting the data, but through the scaling relation $D(2-\tau)=2$. The estimates
for $\fssExpo{s}{1}$ verify this scaling relation. The estimates in the last three columns
should coincide under the narrow-joint-distribution assumption, \Eref{narrow_joint}. Estimates 
for different observables are \emph{not} independent.}
\begin{tabular}{l|lllllllllll}
lattice & $d$ 	& $D$ 		& $\tau$ 	& $z$ 		& $\alpha$ 	& $D_a$ 	& $\tau_a$ 	& $\fssExpo{s}{1}$ & $-\Sigma_s$& $-\Sigma_t$ & $-\Sigma_a$ \\
\hline
\hline
Simple chain\xspace 	& 1 	& 2.27(2) 	& 1.117(8) 	& 1.450(12) 	& 1.19(2) 	& 0.998(4) 	& 1.260(13) 	& 2.000(4) 	& 0.27(2) 	& 0.27(3)	& 0.259(14) \\
Rope ladder\xspace 	& 1 	& 2.24(2) 	& 1.108(9) 	& 1.44(2) 	& 1.18(3) 	& 0.998(7) 	& 1.26(2) 	& 1.989(5) 	& 0.24(2) 	& 0.26(5)	& 0.26(2) \\
\nNNNshort 	& 1 	& 2.33(11) 	& 1.14(4) 	& 1.48(11) 	& 1.22(14) 	& 0.997(15) 	& 1.27(5) 	& 1.991(11) 	& 0.33(11) 	& 0.3(2)	& 0.27(5) \\
\nFUTAshort 	& 1 	& 2.24(3) 	& 1.105(14) 	& 1.43(3) 	& 1.16(6) 	& 0.999(15) 	& 1.24(5) 	& 2.008(11) 	& 0.24(3) 	& 0.23(9)	& 0.24(5) \\
\hline
Square\xspace 	& 2 	& 2.748(13) 	& 1.272(3) 	& 1.52(2) 	& 1.48(2) 	& 1.992(8) 	& 1.380(8) 	& 1.9975(11) 	& 0.748(13) 	& 0.73(4)	& 0.76(2) \\
Jagged\xspace 	& 2 	& 2.764(15) 	& 1.276(4) 	& 1.54(2) 	& 1.49(3) 	& 1.995(7) 	& 1.384(8) 	& 2.0007(12) 	& 0.764(15) 	& 0.76(5)	& 0.77(2) \\
\nARCHshort 	& 2 	& 2.76(2) 	& 1.275(6) 	& 1.54(3) 	& 1.50(3) 	& 1.997(10) 	& 1.382(11) 	& 2.001(2) 	& 0.76(2) 	& 0.78(6)	& 0.76(3) \\
\nNOCRshort 	& 2 	& 2.750(14) 	& 1.273(4) 	& 1.53(2) 	& 1.49(2) 	& 1.992(7) 	& 1.381(8) 	& 2.0005(12) 	& 0.750(14) 	& 0.75(4)	& 0.76(2) \\
Triangular\xspace 	& 2 	& 2.76(2) 	& 1.275(5) 	& 1.51(2) 	& 1.47(3) 	& 2.003(11) 	& 1.388(12) 	& 1.997(2) 	& 0.76(2) 	& 0.71(6)	& 0.78(3) \\
\nKAGOshort 	& 2 	& 2.741(13) 	& 1.270(4) 	& 1.53(2) 	& 1.49(2) 	& 1.993(8) 	& 1.381(9) 	& 1.9994(12) 	& 0.741(13) 	& 0.75(5)	& 0.76(2) \\
Honeycomb\xspace 	& 2 	& 2.73(2) 	& 1.268(6) 	& 1.55(4) 	& 1.51(4) 	& 1.990(13) 	& 1.376(14) 	& 2.000(2) 	& 0.73(2) 	& 0.79(8)	& 0.75(3) \\
\nMITSshort 	& 2 	& 2.75(2) 	& 1.273(6) 	& 1.54(3) 	& 1.50(4) 	& 1.999(12) 	& 1.387(12) 	& 1.998(2) 	& 0.75(2) 	& 0.77(7)	& 0.77(3) \\

\hline
\end{tabular}
\end{table*}

\begin{table*}[t]
\caption{
\tlabel{Exponent_summary}
Overall estimates of scaling exponents in one and two dimensions.
Only the fits using \Eref{fit_TwoQuarter}, based on the data in \Tref{avalanche_exponents} and shown in bold, are fully reliable. Entries for \Eref{fit_OneQuarter} and 
\Eref{fit_D} are for comparison to other estimates only. 
Fits with a goodness of less than $0.1$ are marked by $\qoflow{\cdot}$.
The estimate for $\Sigma$, \Eref{narrow_joint},
is based on all estimates for $D(\tau-1)$, $z(\alpha-1)$ and $D_a(\tau_a-1)$ in \Tref{avalanche_exponents}. Their correlation is 
taken into account by multiplying their respective error by $\sqrt{3}$.  }
\begin{tabular}{lllllllll}
$d$ & function & $D$ 	& 	$\tau$ & 	$z$ 	& 	$\alpha$ & 	$D_a$ 		& $\tau_a$	& $-\Sigma$ \\
\hline
\hline
\textbf{1} &	\textbf{\Eref{fit_TwoQuarter}}	&	 \textbf{2.253(14)}	&	\textbf{1.112(6)}	&	\textbf{1.445(10)}	&	\textbf{1.18(2)}	&	\textbf{0.998(3)}	&	\textbf{1.259(11)}	&	\textbf{0.26(2)} \\
1 &	\Eref{fit_OneQuarter}	&	 \qoflow{2.265(4)}	&	\qoflow{1.117(2)}	&	\qoflow{1.449(2)}	&	1.172(3)	&	1.0000(6)	&	1.249(2)	&	0.249(3) \\
1 &	\Eref{fit_D}	&	 \qoflow{2.2520(3)}	&	\qoflow{1.11188(11)}	&	\qoflow{1.4632(6)}	&	\qoflow{1.219(2)}	&	1.0000(8)	&	1.276(2)	&	\qoflow{0.297(3)} \\
\hline
\textbf{2} &	\textbf{\Eref{fit_TwoQuarter}}	&	 \textbf{2.750(6)}	&	\textbf{1.273(2)}	&	\textbf{1.532(8)}	&	\textbf{1.4896(96)}	&	\textbf{1.995(3)}	&	\textbf{1.382(3)}	&	\textbf{0.761(13)} \\
2 &	\Eref{fit_OneQuarter}	&	 2.7698(12)	&	1.2779(3)	&	1.5407(14)	&	1.498(2)	&	1.9990(5)	&	1.3843(6)	&	0.768(2) \\
2 &	\Eref{fit_D}	&	 \qoflow{2.7673(3)}	&	\qoflow{1.27728(7)}	&	1.541(2)	&	1.501(2)	&	1.9985(6)	&	1.3853(6)	&	0.770(2) \\

\hline
\end{tabular}
\end{table*}

\subsection{Moment ratios}
Similar to the plain moments, one has to allow for corrections when fitting
moment ratios. Most two-dimensional lattices (except \NOCR and \MITS) produced
consistent results with a goodness of fit of greater than $0.1$ with
\begin{equation}
\momrat{x}{n} + D_{x,n} N^{-0.25}
\elabel{momrat_OneQuarter}
\end{equation}
but in order to capture all lattices and for consistency with the above we decided to add a further correction, 
finally fitting against
\begin{equation}
\momrat{x}{n} + D_{x,n} N^{-0.25} + E_{x,n} N^{-0.5} \ .
\elabel{momrat_TwoQuarter}
\end{equation}
In contrast to the finite size scaling exponents $\fssExpo{x}{n}$ of the
moments considered above, all moment ratios were fitted as if they were
independent, \ie considering them simultaneously may be misleading as they are
correlated and these correlations have not been accounted for.

The results are shown in \Tref{momrats_list}. In general, two dimensional lattices
are much better than one dimensional ones. The observable most easily fitted is
the area size distribution (which might be caused by the plain amplitudes
$A_{a,n}$ being universal, see below). The only two-dimensional lattice that
displayed low goodness of fit throughout was the \nTRIA, while the \nHONE had a
single poorly fitting ratio. In one dimension, the picture is reversed; There is hardly
any reasonably fittable moment ratio. Fits which produced a goodness of less than
$0.1$ are marked in \Tref{momrats_list} again by $\qoflow{\cdot}$. Results are
rather noisy for the highest moment ratios, which might suggest an explanation for
the slight inconsistencies, which are not covered by the statistical error, for example
for $\momrat{a}{6}$ in \NOCR and \MITS. Yet, in one dimension, it is the lower
order moment ratios that were most difficult to handle.

\begin{table*}[t]
\caption{\tlabel{momrats_list}
Estimates for the moment ratios as defined in \Eref{moment_ratio} obtained
by fitting the relevant ratios against \Eref{momrat_TwoQuarter}.
Fits with a goodness of less than $0.1$ are marked by $\qoflow{\cdot}$.
}
\begin{tabular}{lllllll}
lattice	&$d$	& $x$	&$\momrat{x}{3}$	& $\momrat{x}{4}$	& $\momrat{x}{5}$	& $\momrat{x}{6}$ \\
\hline\hline
Simple chain\xspace 		& 1 	& $s$	& \qoflow{1.482(4)}	& \qoflow{2.68(2)}	& \qoflow{5.40(11)}	& \qoflow{11.3(5)}	\\
Rope ladder\xspace 		& 1 	& $s$	& \qoflow{1.453(6)}	& \qoflow{2.48(3)}	& \qoflow{4.43(15)}	& \qoflow{7.2(6)}	\\
\nNNNshort 		& 1 	& $s$	& \qoflow{1.479(13)}	& \qoflow{2.61(7)}	& \qoflow{5.0(3)}	& \qoflow{9.4(13)}	\\
\nFUTAshort 		& 1 	& $s$	& \qoflow{1.412(10)}	& \qoflow{2.26(6)}	& \qoflow{3.5(2)}	& \qoflow{3.58(97)}	\\
\hline
Square\xspace 		& 2 	& $s$	& 1.825(3)	& 4.35(2)	& 12.25(14)	& 39.0(8)	\\
Jagged\xspace 		& 2 	& $s$	& 1.830(3)	& 4.38(2)	& 12.44(14)	& 39.9(8)	\\
\nARCHshort 		& 2 	& $s$	& 1.821(4)	& 4.32(3)	& 12.1(2)	& 37.86(99)	\\
\nNOCRshort 		& 2 	& $s$	& 1.828(3)	& 4.36(2)	& 12.30(14)	& 39.2(8)	\\
Triangular\xspace 		& 2 	& $s$	& \qoflow{1.830(5)}	& \qoflow{4.37(3)}	& \qoflow{12.3(2)}	& \qoflow{39.1(11)}	\\
\nKAGOshort 		& 2 	& $s$	& 1.832(3)	& 4.40(3)	& 12.6(2)	& 40.9(9)	\\
Honeycomb\xspace 		& 2 	& $s$	& 1.829(5)	& 4.38(4)	& 12.5(2)	& 40.2(11)	\\
\nMITSshort 		& 2 	& $s$	& 1.820(5)	& 4.31(4)	& 12.0(2)	& 37.4(11)	\\
\hline\hline
Simple chain\xspace 		& 1 	& $t$	& \qoflow{1.472(3)}	& \qoflow{2.63(2)}	& \qoflow{5.34(7)}	& \qoflow{11.9(3)}	\\
Rope ladder\xspace 		& 1 	& $t$	& \qoflow{1.455(4)}	& \qoflow{2.52(2)}	& \qoflow{4.84(9)}	& \qoflow{9.8(4)}	\\
\nNNNshort 		& 1 	& $t$	& \qoflow{1.470(9)}	& \qoflow{2.58(5)}	& \qoflow{5.0(2)}	& 10.3(8)	\\
\nFUTAshort 		& 1 	& $t$	& \qoflow{1.437(6)}	& \qoflow{2.43(3)}	& \qoflow{4.44(13)}	& \qoflow{8.2(5)}	\\
\hline
Square\xspace 		& 2 	& $t$	& 2.116(2)	& 5.95(2)	& 19.90(13)	& 75.7(9)	\\
Jagged\xspace 		& 2 	& $t$	& 2.117(2)	& 5.96(2)	& 20.06(13)	& 77.0(9)	\\
\nARCHshort 		& 2 	& $t$	& 2.115(3)	& 5.94(2)	& 19.9(2)	& 76.1(11)	\\
\nNOCRshort 		& 2 	& $t$	& 2.114(2)	& 5.93(2)	& 19.78(13)	& 74.9(9)	\\
Triangular\xspace 		& 2 	& $t$	& 2.113(4)	& \qoflow{5.93(3)}	& \qoflow{19.8(2)}	& \qoflow{75.0(13)}	\\
\nKAGOshort 		& 2 	& $t$	& 2.116(3)	& 5.96(2)	& 20.04(14)	& 76.95(96)	\\
Honeycomb\xspace 		& 2 	& $t$	& 2.110(4)	& 5.89(3)	& 19.6(2)	& 74.4(12)	\\
\nMITSshort 		& 2 	& $t$	& 2.110(3)	& 5.90(3)	& 19.7(2)	& 74.8(12)	\\
\hline\hline
Simple chain\xspace 		& 1 	& $a$	& 1.3318(6)	& 1.961(2)	& 3.037(5)	& 4.839(11)	\\
Rope ladder\xspace 		& 1 	& $a$	& 1.3301(11)	& 1.957(4)	& 3.029(9)	& 4.83(2)	\\
\nNNNshort 		& 1 	& $a$	& 1.340(2)	& \qoflow{1.990(7)}	& \qoflow{3.11(2)}	& \qoflow{5.00(4)}	\\
\nFUTAshort 		& 1 	& $a$	& 1.332(2)	& 1.962(6)	& 3.038(15)	& 4.85(3)	\\
\hline
Square\xspace 		& 2 	& $a$	& 1.7501(11)	& 3.709(5)	& 8.69(2)	& 21.66(7)	\\
Jagged\xspace 		& 2 	& $a$	& 1.7496(10)	& 3.710(5)	& 8.70(2)	& 21.72(7)	\\
\nARCHshort 		& 2 	& $a$	& 1.7503(14)	& 3.712(7)	& 8.70(3)	& 21.711(98)	\\
\nNOCRshort 		& 2 	& $a$	& 1.7517(10)	& 3.718(5)	& 8.72(2)	& 21.78(7)	\\
Triangular\xspace 		& 2 	& $a$	& 1.749(2)	& 3.710(8)	& 8.70(3)	& 21.73(11)	\\
\nKAGOshort 		& 2 	& $a$	& 1.7500(10)	& 3.713(5)	& 8.71(2)	& 21.78(7)	\\
Honeycomb\xspace 		& 2 	& $a$	& \qoflow{1.747(2)}	& 3.698(9)	& 8.66(3)	& 21.62(11)	\\
\nMITSshort 		& 2 	& $a$	& 1.748(2)	& 3.703(8)	& 8.67(3)	& 21.62(11)	\\

\hline
\end{tabular}
\end{table*}

The overall estimates for the moment ratios are shown in \Tref{generateMomRatTable_summary}. We refrained
from stating an overall estimate, where not at least two lattices produced reliable estimates
as shown in \Tref{momrats_list}. Those that show low quality of fit are marked as above.

There is, again, a certain sensitivity to the fitting function; \Eref{momrat_OneQuarter} gives slightly
incompatible results, with remarkably small error bars. Given what has been said about 
the goodness of fit, we place our confidence in the results presented in \Tref{generateMomRatTable_summary}.

\begin{table}[t]
\caption{
Overall estimates for the moment ratios defined in \Eref{moment_ratio}, based 
on the data presented in \Tref{momrats_list}, if enough data was available.
\tlabel{generateMomRatTable_summary}
}
\begin{tabular}{llllll}

$d$& x	&$\momrat{x}{3}$	& $\momrat{x}{4}$	& $\momrat{x}{5}$	& $\momrat{x}{6}$ \\
\hline\hline
1 &	 s	&	---	&	---	&	---	&	---	\\
1 &	 t	&	---	&	---	&	---	&	---	\\
1 &	 a	&	\qoflow{1.3320(5)}	&	1.961(2)	&	3.035(4)	&	4.838(9)	\\
\hline
2 &	 s	&	1.8273(14)	&	4.363(10)	&	12.32(6)	&	\qoflow{39.3(3)}	\\
2 &	 t	&	2.11423(98)	&	5.939(8)	&	19.88(6)	&	75.8(4)	\\
2 &	 a	&	1.7501(4)	&	3.711(2)	&	8.699(8)	&	21.72(3)	\\
\hline

\hline
\end{tabular}
\end{table}

\section{Discussion}
\slabel{discussion}
Exponents (\Tref{avalanche_exponents}) and, at least in two dimensions, 
moment ratios (\Tref{momrats_list}) are universal across different lattices.
Subleading orders of moments are noisy, but still fairly consistent. There
can be little doubt that the Abelian Manna model displays all the hallmarks of a
critical system as they are known from equilibrium critical phenomena. The
exponents shown in bold in \Tref{Exponent_summary} and the moment ratios in 
\Tref{generateMomRatTable_summary} (except those shown in brackets) are
perfectly consistent across all our results and represent reliable, high accuracy
estimates of the universal quantities characterising this universality class.

\begin{table*}[t]
\caption{Comparison of the results in the present work to the estimates found in the literature.
Many of the works quoted below have studied variants of the Manna Model.
The values taken from
\cite{Christensen:2004,PruessnerPHDThesis:2004} and \cite{Bonachela:2008} in two dimensions are for the Oslo Model.
The exponents marked as DP are those for the directed percolation universality class. They are 
derived via scaling laws \cite{Luebeck:2004}.
\tlabel{literature_comparison}
}
\begin{tabular}{lllllllll}
$d$ & reference				& $D$ 		& $\tau$ 	& $z$ 		& $\alpha$ 	& $D_a$ 	& $\tau_a$	& $-\Sigma$ \\
\hline
\hline
1 & this work			  	& $2.253(14)$   & $1.112(6)$    & $1.445(10)$	& $1.18(2)$     & $0.998(3)$    & $1.259(11)$   & $0.26(2)$ \\
1 & \cite{NakanishiSneppen:1997} 	& $2.2(1)$ 	& $1.09(3)$ 	& $1.47(7)$ 	&		&		&		&	\\
1 & \cite{DickmanCampelo:2003}		& 		& $1.11(2)$	& 		& $1.18(2)$	&		&		&	\\
1 & \cite{LuebeckHeger:2003a}		& 		& $1.11(2)$	& 		& $1.17(3)$	&		&		&	\\
1 & \cite{Bonachela:2008}		& 		& $1.11(5)$	& 		& $1.17(5)$	&		&		&	\\
1 & \cite{Christensen:2004}		& $2.2496(12)$	& 		&		&		&		&		&	\\
1 & \cite{PruessnerPHDThesis:2004}	& $2.2509(6)$ 	&		&		&		&		&		&	\\
\hline
1 & \cite{Jensen:1999} (DP)		& $2.328673(12)$&		&$1.580745(10)$ &		& $1$		&		&	\\
\hline
2 & this work				& $2.750(6)$	& $1.273(2)$	& $1.532(8)$	& $1.4896(96)$  & $1.995(3)$	& $1.382(3)$	& $0.761(13)$ \\
2 & \cite{Manna:1991a}			& $2.75$	& $1.28(2)$	& $1.55$	& $1.47(10)$	&		&		&	\\
2 & \cite{ChessaVespignaniZapperi:1999} & $2.73(2)$	& $1.27(1)$	& $1.50(2)$	& $1.50(1)$	& $2.02(2)$	& $1.35(1)$	&	\\
2 & \cite{Luebeck:2000}			& $2.76(1)$	& 		& $1.54(1)$	& 		& $2.03(1)$	&		& $0.75(4)$ \\
2 & \cite{DickmanCampelo:2003}		&		& $1.30(1)$	&		& $1.55(4)$	&		&		&	\\
2 & \cite{LuebeckHeger:2003a}		& 		& $1.28(14)$	&		& $1.50(3)$	&		&		&	\\
2 & \cite{Bonachela:2008}		& 		& $1.26(3)$	&		& $1.48(3)$	&		&		&	\\
\hline
2 & \cite{VoigtZiff:1997} (DP)		& $2.979(2)$	&		& $1.765(3)$	&		& $2$		&		&	\\
\hline
\end{tabular}
\end{table*}

In the literature can be found
a large number of estimates for the exponents in \Tref{Exponent_summary}. Traditionally,
the AMM is studied in two dimensions and therefore many more estimates are available in 
that dimension. Given that the AMM is in the same universality class as the Oslo Model
\cite{ChristensenETAL:1996,NakanishiSneppen:1997}, there is a second source for
comparison. What makes the comparison more complicated is the fact that many authors
have studied variants of the Manna Model (in fact, the Abelian version studied here
is a variant of the original model); for example Dickman and Campelo studied
the Manna Model with height restrictions \cite{DickmanCampelo:2003} and
L{\"u}beck and Heger studied its ``fixed energy sandpile'' version \cite{LuebeckHeger:2003a,VespignaniETAL:2000}. 
\Tref{literature_comparison} collects a broad range of estimates across the literature, 
which nevertheless provide a perfectly consistent picture. The present work clearly improves on 
comprehensiveness and on accuracy, which for most estimates is improved by one digit. 
The fact that some estimates in the literature have even smaller error bars than 
ours might be partly due to our over-estimation of statistical errors
but also due to other authors using models that are better behaved in one dimension, as in 
\cite{Christensen:2004,PruessnerPHDThesis:2004}.

\Tref{literature_comparison} also contains the avalanche exponents for the directed percolation (DP) 
universality class. The exponent $D$ is derived through the identity $D=d+z-\beta/\nu_{\perp}$, where 
$-\beta/\nu_{\perp}$ is the finite size scaling exponent of the activity in DP \footnote{Voigt and
Ziff follow a different notation; $2/z$ in their work \cite{VoigtZiff:1997} corresponds to $z$ here 
and their $2\eta/z$ to $\beta/\nu_{\perp}$ here.}. The dynamical exponent $z$ is sometimes used 
in the DP literature as what would have been in the present notation $2/z$.
The exponent for $D_a$ is
based on the asumption of compact avalanches \cite{Luebeck:2004}.
Although sometimes disputed in the literature 
\cite{BonachelaMunoz:2007,BonachelaMunoz:2008}
using avalanche exponents leaves little doubt that the Manna universality class differs from DP.
As DP is normally performed with periodic boundaries and with different observables, there is,
to our knowledge
unfortunately no published work on the moment ratios we considered here. In fact, depending 
on the definition of the ensemble \cite{MarroDickman:1999,LuebeckHeger:2003a,Pruessner:2007b,Pruessner:2008b}, 
some moment ratios in DP can be undefined.

Certain finite size scaling features specific to SOC are confirmed as well: 
Compactness of avalanches, $D_a=d$ is very strongly supported (in two dimensions
the deviation of $D_a$ from $d$ is slightly bigger than one standard deviation, though), as is the
universality of $\Sigma$, \Eref{def_Sigma}. It is reassuring that the asymptotics
of the first moment of the avalanche size, $\fssExpo{s}{1}=2$, are recovered,
validating our numerical schemes.

Although we invested more than twice as much CPU time (in absolute terms) in
one dimensional lattices, the results are significantly noisier than for two
dimensional lattices. Providing a sufficient number of correction terms still produces
consistent data (\Tref{Exponent_summary}), but the results for two dimensional
lattices are clearly superior. In fact, the error bar on the exponents for two
dimensional lattices is typically half that of one dimensional lattices.

It is known that the Manna Model suffers from significant logarithmic
corrections \cite{DickmanCampelo:2003}. Manna himself noted a ``considerable
curvature'' in what should have been a straight line in a double logarithmic
plot \cite{Manna:1991a}. Similarly, L{\"u}beck and Heger
\cite{LuebeckHeger:2003a} found a surprising splitting of the Manna universality 
class in one dimension, which might also be due to the presence of significant 
corrections.

The results for the finite size scaling exponents in \Tref{Exponent_summary} suggest that one-dimensional systems
are more difficult to fit, with the alternative fitting functions
\Eref{fit_OneQuarter} and \Eref{fit_D} both clearly performing worse than in two
dimensions. One might think that some of the problems are caused by having much
higher accuracy in the estimates in one dimension (and thus requiring more correction
term, as bad choices for the fitting function can no longer be hidden in a large 
statistical
error), given that we spent, per lattice, typically about five
times more CPU time than in two dimensions. The opposite is the case (probably because 
correlation times grow like a power law of the linear extent which are very large in 
one dimension): Depending on the observable, relative errors are between a factor $2$
and $10$ worse in \nLINE compared to \nSQUA. This holds similarly for other lattices, 
except for the \nNNN, on which we spent
less than $1/6$ of the CPU time we spent on the other one dimensional lattices.

In general
the relative statistical error vanishes like the inverse square of the CPU time, 
so that the product of the two gives a measure of ``efficiency'' of a lattice.
Using that measure, the \nLINE is the most efficient, followed by \nNNN, \nLADD
and finally the \nFUTA, which is about a factor $4$ less efficient ($4$ times
the CPU time is needed for results with similar relative error). This statement, 
however, is put in perspective, by noting that we used variety of different hardware
throughout. The two-dimensional lattices fall roughly in three classes: \nNOCR, \nKAGO,
\nJAGG, \nSQUA, followed by \nTRIA, \nARCH and \nMITS, and finally the \nHONE. The latter
is clearly the worse (again by about a factor $4$), while the \nTRIA is in the first
group for some of the observables.

There is a caveat, however. Within a given amount of CPU time and for a given system size, 
the \NOCR produces a larger number of avalanches, which are
typically much smaller than those for other lattices, because the fraction
of virtual neighbours is about $1.5$ times higher for the \nNOCR than
for, say, the \nHONE, so that particles are dissipated more frequently.
As a result, statistics are 
comparatively better for \nNOCR, which, however, may pose higher demands
on correction terms with generally larger amplitudes. We could, however,
not identify a systematic behaviour in this respect. For example, the moments of
\nJAGG are, within error, the same as for $\nSQUA$, yet the latter has a
much smaller average fraction of dissipative links. In fact, as discussed 
below, the same 
leading order amplitudes are found for the avalanche area distribution across
all lattices of the same dimension.

The one dimensional lattice perform particularly badly for the moment ratios. Essentially only those
for the area size distribution can be fitted well and then produce fairly consistent 
results (except for the \nNNN). Originally we expected improved scaling behaviour with the 
introduction of next nearest neighbour interaction, as it prevents degeneracy issues (and,
say conserved quantities) as they are sometimes observed on the \nLINE \cite{HughesPaczuski:2002,DharRamaswamy:1989}.
The two dimensional lattices, with the exception of the
\nTRIA, generally behave much better, when fitting moment ratios. Given the importance of the boundary conditions, 
it is remarkable how well all universal
quantities addressed in the present work are reproduced, even when
some two dimensional lattices like the \KAGO and the \JAGG have rather complicated 
boundaries (although the latter is the \nSQUA in the thermodynamic limit) and many
of the lattices have an aspect ratio of unity only asymptotically.

To our surprise, the amplitudes $A_{a,n}$ in \Eref{fit_TwoQuarter} 
obtained when fitting the moments of the area distribution
seem to be universal themselves. According to \Eref{xn_asymptote}
\begin{equation}
A_{a,n} = a_a b_a^{n+1-\tau_a}
\int_0^\infty \dint{y} y^{n-\tau_x} \GC_a(y) \ ,
\end{equation}
which is universal provided the metric factors $a_a$ and $b_a$ are,
which is not normally the case. Since $D_a=d$, however, the amplitude
$b_a$ of the cutoff $b_a L^{D_a}$ is dimensionless. It is therefore 
reasonable to assume that it is universal. There is no reason, however,
to assume that the same should hold for $a_a$ --- the argument that $a_a$ is
determined by normalisation does not hold as \Eref{power_law_distribution}
applies only asymptotically, \ie the fraction of small event sizes which do not
follow simple scaling can, in principle, vary from lattice to lattice. Moreover,
$a_a$ is dimensionful since $\tau_a\ne1$. Nonetheless,
it turns out that it does not vary.
As a result,
to leading order, the moments of the avalanche areas 
in one dimension follow
\begin{eqnarray}
\ave{a^1} &=& \qoflow{2.01(7)} N^{2-1.259(11)}\\
\ave{a^2} &=& \qoflow{1.21(5)} N^{3-1.259(11)}\\
\ave{a^3} &=& \qoflow{0.96(4)} N^{4-1.259(11)}\\
\ave{a^4} &=& \qoflow{0.83(4)} N^{5-1.259(11)}\\
\ave{a^5} &=& \qoflow{0.76(4)} N^{6-1.259(11)}
\elabel{area_amplitudes_one_dim}
\end{eqnarray}
and in two dimensions
\begin{eqnarray}
\ave{a^1} &=& \qoflow{0.756(7)} N^{2-1.382(3)}\\
\ave{a^2} &=& 0.217(3) N^{3-1.382(3)}\\
\ave{a^3} &=& 0.109(3) N^{4-1.382(3)}\\
\ave{a^4} &=& 0.066(3) N^{5-1.382(3)}\\
\ave{a^5} &=& 0.045(2) N^{6-1.382(3)}
\end{eqnarray}
as a function of the number $N$ of sites, independent of the lattice type.
We need to qualify the statement, by pointing out that
in one dimension the amplitudes across different lattices are rather noisy, and in two dimension
the amplitude of $\ave{a^1}$ has a goodness of fit of just under $0.1$ 
(we are using the $\qoflow{\cdot}$ notation again here), while the other
results in two dimensions all have a goodness of fit better than $0.5$.

Remarkably,
had we fitted the area moments against a lattice-dependent multiple 
of $N^{1/2}$, such as the perceived linear extent of the lattice,
then this multiplier would have shown in the resulting amplitude $A_{a,n}$.
and so the apparently universal behaviour would not have come to light.

As explained above, the fitting function is a \emph{hypothesis} and ultimately
has an impact on the results. As the number of free parameters increases so does
the susceptibility of the result on the initial condition. The approach described
above, using fairly large systems (with weaker corrections) with not too small
error bars, in conjunction with simple fitting functions, the initial values of
which are determined by those with fewer terms, seems to produce robust and reliable
results. Comparing the results based on \Eref{fit_TwoQuarter} and \Eref{fit_OneQuarter}
in \Tref{Exponent_summary} to those on the basis of \Eref{fit_D} indicates that the
former are superior. An acceptable goodness of fit is reached for \Eref{fit_D} only 
for those exponents that coincide within a bit more than one standard deviation
with the estimates based on \Eref{fit_TwoQuarter}.
\Eref{fit_OneQuarter}, on the other hand, in summary (\Tref{Exponent_summary}) 
coincides with \Eref{fit_TwoQuarter}, but some, individual finite size scaling exponents,
such as $\fssExpo{s}{1}$, but also $\Sigma_x$ and $D_a$, were estimated too poorly.

In summary, the present work confirms the Abelian Manna Model as an SOC model
that displays non-trivial, robust, reproducible, universal scaling behaviour 
in one and two dimension across different lattices.

\begin{acknowledgments}
H. N. Huynh gratefully acknowledges the hospitality of Department of Mathematics,
Imperial College London and GP that of the Department of
Physics at NTU Singapore.
The authors are indebted to Andy
Thomas, Dan Moore and Niall Adams for running the SCAN computing facility at the
Department of Mathematics of Imperial College London.
\end{acknowledgments}

\bibliography{bib_extracted}
\end{document}